\documentclass[aps,prd,preprint,eqsecnum,nofootinbib,showkeys,showpacs,groupedaddress,a4paper,byrevtex]{revtex4-1}
\usepackage{graphicx, subfigure}
 \usepackage{amsthm,amssymb,amsmath,amsfonts,bm}

\DeclareMathOperator{\sign}{ sgn}
\begin{document}

\title{Three two-component fermions with contact interactions:
correct formulation and energy spectrum}

\author{O.~I.~Kartavtsev}\email{oik@nusun.jinr.ru}
\author{A.~V.~Malykh}\email{maw@theor.jinr.ru}
\affiliation{Joint Institute for Nuclear Research, Dubna, 141980, Russia} 

\date{\today}

\begin{abstract} 
Properties of two identical particles of mass $m$ and a distinct 
particle of mass $m_1$ in the universal low-energy limit of zero-range 
two-body interaction are studied in different sectors of total angular 
momentum $L$ and parity $P$. 
For the unambiguous formulation of the problem in the interval 
$\mu_r(L^P) < m/m_1 \le \mu_c(L^P)$ ($\mu_r(1^-) \approx 8.619$ and 
$\mu_c(1^-) \approx 13.607$, $\mu_r(2^+) \approx 32.948$ and 
$\mu_c(2^+) \approx 38.630$,~etc.) in each $L^P$ sector
an additional parameter $b$ determining the wave function near 
the triple-collision point is introduced; thus, a one-parameter family 
of self-adjoint Hamiltonians is defined. 
Within the framework of this formulation, dependence of the bound-state 
energies on $m/m_1$ and $b$ in the sector of angular momentum and 
parity $L^P$ is calculated for $L \le 5$ and analysed with the aid 
of a simple model. 
A number of the bound states for each $L^P$ sector is analysed and 
presented in the form of ``phase diagrams'' in the plane of two 
parameters $m/m_1$ and $b$.

\end{abstract}


\maketitle

\section{Introduction.} 
\subsection*{Motivation}
In the present time, properties of multi-component ultra-cold quantum 
gases are on demand experimentally~\cite{Jag16} and 
theoretically~\cite{Ospelkaus06,Fratini12,Iskin06,Levinsen09,Mathy11,Alzetto12,Jag14}. 
Different aspects of few-body dynamics in two-species mixtures has 
attracted much attention. 

At low energy the dependence on potential form is disappear, therefore, 
zero-range model (ZRM) is a good approximation for such systems. 
There are many advantages of using ZRM. 
First at all, the only parameter of the interaction, namely, 
the two-body scattering length $a$, can be taken as a scale, (or scale 
$a\to\infty$) that lead to parameterless  description of the two-body  
problem. 
Then, for three- and more-body problem, one expects the few-parameter 
or even the parameterless description of the essential dynamical 
features of such systems. 
The usage of ZRM allows one to obtain simple and  even exact 
solutions or reduce the calculation problems with increasing a number 
of particle in system, obtain the predictions of new effects and make 
some proposals for the future study most interesting one theoretically 
or experimentally. 
Moreover, model provides full description within limited class and 
allows to calculate universal constants (such as energies of the bound  
states, scattering characteristics, critical parameters of the system 
and so one).

\subsection*{Problem}
The present paper is devoted to one of the principal issue, the study 
of few two-species particles, namely, two identical particles (bosons or 
fermions) of mass $m$ interacting with a distinct particle of mass $m_1$
in the s-wave. 
In the universal low-energy limit, the interaction between two identical 
fermions is forbidden in the s-wave and is strongly suppressed 
between two heavy bosons in the states of $L > 0$, so that explain why one 
neglect the interaction of identical particles.
To obtain the universal (independent of the particular form of the 
interaction) description of the system, the two-body interaction is taken 
in the framework of the ZRM. 
Then, by using proper units, one could expect formally the one-parameter 
$m/m_1$-dependence of the few-body properties.

\subsection*{Former results}

One of the main features of this three-body problem (namely, two identical 
fermions and a distinct particle) is a principal role of the states with 
unit total angular momentum and negative parity $L^P = 1^-$ in 
the low-energy processes~\cite{Kartavtsev07,Petrov03,Kartavtsev16}. 
As it was already pointed out, there are three regions of the mass ratios,
$m/m_1 < \mu_r$, $\mu_r < m/m_1 \le \mu_c$, and $m/m_1 > \mu_c$, where 
$\mu_c \approx 13.607$, $\mu_r\approx 8.619$.
In the first region, formal three-body Hamiltonian is self-consistent and 
there is zero and one bound state for $m/m_1<8.17$ and $8.17\le m/m_1\le 
\mu_r$, respectively~\cite{Kartavtsev07}. 
For other two regions, the formal construction of the Hamiltonian does 
not obviously provide an unambiguous definition of the three-body problem; 
in particular, one is required an additional parameter, which determines 
the wave function in the vicinity of the triple-collision point (TCP). 
The third region, $m/m_1 >\mu_c$, is well-known Efimov one with 
an infinite number of the bound states~\cite{Efimov73}. 
The necessity of correct formulation of the three-body problem for 
the second region, $\mu_r < m/m_1 \le \mu_c$, was indicated in both 
physical~\cite{Nishida08,Safavi-Naini13,Kartavtsev14} and 
mathematical~\cite{Minlos11a,Minlos12,Minlos14,Minlos14a,Correggi12,
Correggi15,Correggi15a} papers. 
As was done  in~\cite{Kartavtsev16}, a one-parameter family of 
self-adjoint Hamiltonians was defined by introducing an additional 
three-body parameter $b$, which has a meaning of three-body scattering 
length.
As a result, the properties of the energy spectrum of three-body system 
with two identical fermions for $L^P=1^-$ sector is studied in dependence 
on the mass-ratio and parameter $b$, whereas the scattering properties was 
investigated in dependence on the mass-ratio for particular case of 
the three-body parameter $b = 0$~\cite{Kartavtsev07,Petrov03}. 

Despite this, one have to extend the problem for the arbitrary $L^P$ 
sector of angular momentum $L$ and parity $P$ and to consider 
simultaneously the system with two identical bosons and a distinct 
particle.
As it is already known, the bound states can be found only for odd $L$ and 
$P$ if identical particles are fermions and for even $L$ and $P$ if 
identical particles are bosons.
Such systems will be considered below.
As in $L^P = 1^-$ sector, there are three regions of the mass ratios,
$m/m_1 \le \mu_r(L^P)$, $\mu_r(L^P) < m/m_1 \le \mu_c(L^P)$, and 
$m/m_1 >\mu_c(L^P)$, where values $\mu_r(L^P)$ and 
$\mu_c(L^P)$ are presented in Table~\ref{tab1}. 
As follows from the analyses of the wave function in the vicinity of 
the TCP~\cite{Petrov03,Nishida08,Kartavtsev14,Kartavtsev16}, the problem 
of the correct formulation exists and has been done for mass ratio values 
$\mu_r(L^P) < m/m_1 \le \mu_c(L^P)$. 
Until now, in a number of reliable investigations of three two-component 
particles (for $m/m_1 \le \mu_c(L^P)$)~\cite{Petrov03,Kartavtsev07,Kartavtsev07a,Endo11,Helfrich11} 
it was explicitly or implicitly assumed the fastest decrease of the wave 
function near the TCP,~i.e., that correspond to $b = 0$. 

\subsubsection*{Goal}
The main aim of this paper is to formulate the three-body problem 
in the mass-ratio region $\mu_r(L^P) < m/m_1 \le \mu_c(L^P)$ in 
an arbitrary $L^P$ sector (more exactly, for odd $L$ and $P$ if identical 
particles are fermions and for even $L$ and $P$ if identical particles 
are bosons) by introducing the additional parameter $b$ as it was done 
in~\cite{Kartavtsev16} for $L^P = 1^-$ sector. 
In such way, one has to construct a family of self-adjoint 
Hamiltonians, which depends on one parameter $b$, describing 
the solution behaviour at the TCP. 
Then one need to calculate the energy spectrum in an arbitrary $L^P$ 
sector as a function of $m/m_1$ and a set of three-body parameters 
$\{b_L\}$. 

Consider a three-body problem for particle $1$ of mass $m_1$ 
interacting with two identical particles $2$ and $3$ of mass 
$m_2 = m_3 = m$ via the zero-range potential, which is completely 
described by the scattering length $a$. 
In the center-of-mass frame, define the scaled Jacobi variables as 
${\mathbf x} = \displaystyle\sqrt{2\mu}\left({\mathbf r}_2 - 
{\mathbf r}_1\right)$ and 
${\mathbf y} = \displaystyle\sqrt{2\tilde\mu}\left({\mathbf r}_3 - 
\frac{m_1{\mathbf r}_1 + m{\mathbf r}_2}{m_1 + m}\right)$, where 
${\mathbf r}_i$ are the position vectors and 
$\mu = \displaystyle\frac{mm_1}{m + m_1}$, $\tilde{\mu} = 
\displaystyle\frac{m(m + m_1)}{m_1 + 2m}$ are the reduced masses. 
Throughout the paper units are chosen to provide 
$\hbar = |a| = 2m/(1 + m/m_1) = 1$ that gives unit binding 
energies of the two-body subsystems, 
$\varepsilon_{12} = \varepsilon_{13} = 1$. 

The three-body wave function is a solution of 
\begin{equation}
\label{shred_xy}
\left[\Delta_{{\bf x}} + \Delta_{{\bf y}} + E\right] 
\Psi ({\bf x}, {\bf y}) = 0 \ , 
\end{equation} 
and
\begin{equation}
\label{bound1}
\lim_{r \rightarrow 0} \frac{\partial \ln (r \Psi )}{\partial r} 
= -\sign (a) \ , 
\end{equation} 
where the boundary conditions~(\ref{bound1}) with 
$r = |{\mathbf r}_1 - {\mathbf r}_i| $ and $i = 2, 3$ represent 
the zero-range potential in both pairs of distinct particles. 
The Hamiltonian is formally defined by  
Eqs.~(\ref{shred_xy}),~(\ref{bound1}) and depends only on the mass 
ratio $m/m_1$. 
The wave function is symmetrical or anti-symmetrical under permutation 
of identical particles ${\mathrm P}_{23}$, satisfying the condition 
\begin{equation}
\label{symmetry}
{\mathrm P}_{23}\Psi({\mathbf x}, {\mathbf y}) = 
S \ \Psi({\mathbf x}, {\mathbf y}),
\end{equation} 
where $S = -1$ ($S = 1$) indicates that particles $2$ and $3$ are 
fermions (bosons). 
Total angular momentum $L$, its projection $M$, parity $P$, and index 
of permutational symmetry $S$ are conserved quantum numbers, which will 
be used to label the solutions. 
Only the states of parity $P = (-)^L$, i.~e., for $P$ and $L$ 
odd or even simultaneously, need to be considered, as the states 
of opposite parity correspond to three non-interacting particles 
for the zero-range potential. 
As the system's properties are independent of $M$, a complete 
description of the three-body problem, e.~g., energy levels, will be  
given by the formal one-parameter Hamiltonian depending on $m/m_1$ 
in different $\{S, L\}$ sectors. 

\subsection{Hyper-radial equations}
\label{Hyper-radial_equations} 

Let us define a hyper-radius $\rho$ and hyper-angular variables 
$\{ \alpha, \hat{\mathbf x}, \hat{\mathbf y} \}$ by 
$x = \rho\cos\alpha$, $y = \rho\sin\alpha$, $\hat{\mathbf x} = 
{\mathbf x}/x$, and $\hat{\mathbf y} = {\mathbf y}/y$. 
To produce a convenient basis for expansion of the total wave function  
defined by Eqs.~(\ref{shred_xy}),~(\ref{bound1}), introduce 
an auxiliary eigenvalue problem on a hyper-sphere (for fixed parameter 
$\rho $)~\cite{Macek68}, 
\begin{equation}
\label{eqonhypershere}
\left[\frac{1}{\sin^2 2\alpha }\left( \sin^2 2 \alpha 
\frac{\partial}{\partial \alpha } \right) + 
\frac{1}{\sin^2\alpha }\Delta_{\hat{\mathbf x}} + 
\frac{1}{\cos^2\alpha }\Delta_{\hat{\mathbf y}} + \gamma^2 (\rho ) - 
4 \right] \Phi(\alpha, \hat{\mathbf x}, \hat{\mathbf y}; \rho) = 0 \ , 
\end{equation} 
\begin{equation}
\label{bch}
\lim_{\alpha \rightarrow \pi/2} 
\frac{\partial\log\left[ (\alpha - \pi/2)\Phi \right] }
{\partial\alpha }  = \rho \, \sign (a) \ ,  
\end{equation} 
along with the symmetry condition~(\ref{symmetry}) for 
$ \Phi (\alpha, \hat{\mathbf x}, \hat{\mathbf y}; \rho ) $. 
Its square-integrable solutions form an infinite set of functions 
$ \Phi_n(\alpha, \hat{\mathbf x}, \hat{\mathbf y}; \rho ) $ 
enumerated by index $1 \le n < \infty $ in ascending order of  
the corresponding eigenvalues $\gamma^2_n(\rho )$. 
 
Besides fermionic (bosonic) symmetry, the functions 
$\Phi (\alpha, \hat{\mathbf x}, \hat{\mathbf y}; \rho)$ inherit 
all the conserved quantum numbers of the total wave function. 
Solution of~(\ref{eqonhypershere}),~(\ref{bch}) 
satisfying~(\ref{symmetry}) will be found in 
the form~\cite{Kartavtsev07,Kartavtsev07a,Kartavtsev16} 
\begin{equation}
\label{Phi}
\Phi(\alpha, \hat{\mathbf x}, \hat{\mathbf y}; \rho ) = 
(1 + S \ {\mathrm P}_{23})
\frac{\varphi^L(\alpha )}{\sin 2\alpha} Y_{LM}(\hat{\mathbf y})\, ,
\end{equation}
where $Y_{LM}(\hat{\mathbf y})$ is the spherical function. 
The action of ${\mathrm P}_{23}$ in terms of the Jacobi variables is 
given by 
\begin{equation}
\label{xy}
{\mathrm P}_{23} \left(
\begin{array}{c}
{\mathbf x} \\ 
{\mathbf y}
\end{array} \right) = 
\left(
\begin{array}{cr}
 -\sin\omega & \cos\omega \\ 
 -\cos\omega & -\sin\omega
\end{array} \right)
\left(
\begin{array}{c}
{\mathbf x} \\ 
{\mathbf y}
\end{array} \right)\ , 
\end{equation} 
where $\omega$ is related to the mass ratio by $\sin\omega = 1/(1 + m_1/m)$.
To impose the boundary condition~(\ref{bch}), one takes the limit 
$x \to 0$ in Eq.~(\ref{xy}) and finds 
${\mathrm P}_{23} \alpha \to \omega $ and 
${\mathrm P}_{23} Y_{LM}(\hat{\mathbf y}) \to
(-1)^L Y_{LM} (\hat{\mathbf y}) $ in the limit $\alpha \to \pi/2$. 
As a result, one comes to the eigenvalue problem 
\begin{equation} 
\label{eqonhyp1}
\left[\frac{d^2}{d \alpha^2} - \frac{L(L + 1)}{\sin^2\alpha}
 + \gamma^2\right]\varphi^L(\alpha) = 0 \ , 
\end{equation} 
and 
\begin{subequations}
\label{bc}
\begin{equation}
\label{bconhyp1}
\varphi^L(0) = 0 \ ,  
\end{equation} 
\begin{equation}
\label{bconhyp}
 \lim_{\alpha\rightarrow \pi/2}
\left(\frac{d}{d \alpha} - \rho\, \sign (a) \right) 
\varphi^L(\alpha) = \frac{2 S(-)^L}{\sin 2\omega} 
\varphi^L(\omega) \, .
\end{equation} 
\end{subequations}
Solution of~(\ref{eqonhyp1}) and~(\ref{bconhyp1}) is discussed in  
Appendix~\ref{Appendix_hypsol}. 
The boundary condition~(\ref{bconhyp}), along 
with~(\ref{varphi}),~(\ref{LegQ0}), and~(\ref{varphi_deriv_pi/2}), 
gives the transcendental equation 
\begin{eqnarray}
\nonumber
\rho \sign(a) 
\Gamma\left(\frac{L + 1 + \gamma }{2} \right) 
\Gamma\left(\frac{L + 1 - \gamma }{2} \right) & & = \\
\label{transeq1}
2 \Gamma\left(\frac{L + \gamma }{2} + 1 \right) 
\Gamma\left(\frac{L - \gamma }{2} + 1 \right) - 
S & & \frac{2^{1 - L} \pi (\sin\omega )^L}
{\sin\gamma \pi \cos\omega } 
\frac{d^L}{d (\cos \omega )^L} \frac{\sin\gamma \omega }{\sin\omega } 
\end{eqnarray}
determining $ \rho \sign(a) $ as an even single-valued 
function of $\gamma $. 
The inverse function is multi-valued, which different branches form 
a set of eigenvalues $\gamma^2_n(\rho )$ and, accordingly, a set 
of $\varphi_n(\rho )$ and 
$\Phi_n(\alpha, \hat{\mathbf x}, \hat{\mathbf y}; \rho) $. 
In particular, the transcendental equation takes the well-known form 
for $ L = 0 $, 
\begin{equation} 
\label{transeql0}
\rho \sign(a) \sin \gamma \frac{\pi }{2} = 
\gamma \cos \gamma \frac{\pi }{2} - 
2 S \frac{\sin \gamma \omega}{\sin 2 \omega } \ . 
\end{equation}
Expansion of the total wave function~\cite{Macek68}, 
\begin{equation}
\label{Psi}
\displaystyle
\Psi = \rho^{-5/2} \sum_{n = 1}^{\infty} f_n(\rho) 
\Phi_n(\alpha, \hat{\mathbf x}, \hat{\mathbf y};\rho)\, ,
\end{equation}
leads to a system of hyper-radial equations (HREs) for the channel 
functions $f_n(\rho)$, 
\begin{equation}
\label{system1}
\left[\frac{d^2}{d \rho^2} - V_n(\rho ) + E \right] f_n(\rho) - 
\sum_{m \neq n}^{\infty}\left[P_{nm}(\rho) - Q_{nm}(\rho)
\frac{d}{d\rho} - \frac{d}{d\rho}Q_{nm}(\rho) \right] f_m(\rho) = 0 \, .
\end{equation}
Here the diagonal terms
\begin{equation}
\label{Vef}
V_n(\rho ) = \frac{\gamma_n^2(\rho) - 1/4}{\rho^2} + P_{nn}(\rho) 
\end{equation} 
play a role of the effective channel potentials, the coupling terms 
are defined as 
$Q_{nm}(\rho) = \displaystyle \left\langle\Phi_n \biggm|
\frac{\partial\Phi_m}{\partial\rho}\right\rangle $ and 
$ P_{nm}(\rho) = \displaystyle
\left\langle\frac{\partial\Phi_n}{\partial\rho} \biggm|
\frac{\partial\Phi_m}{\partial\rho}\right\rangle $, and the notation 
$\langle\cdot|\cdot\rangle$ means integration over the invariant 
volume on a hypersphere 
$\sin^2{2\alpha} \, d\alpha \, d \hat{\mathbf x} d \hat{\mathbf y}$. 
For the zero-range interaction, suitable analytical expressions  
via $\gamma^2_n(\rho )$ and their derivatives are 
derived~\cite{Kartavtsev99,Kartavtsev06,Kartavtsev07}, 
\begin{eqnarray}
\label{Qanal}
& & \hskip -.8cm Q_{nm}(\rho) = 
\left( \gamma_n^2 - \gamma_m^2 \right)^{-1}
\sqrt{\frac{d \gamma_n^2}{d \rho} \frac{d \gamma_m^2}{d \rho}}\, , \\ 
\label{Panal}
& & \hskip -.8cm P_{nm}(\rho) = 
Q_{nm}(\rho) \left[\frac{1}{ \left( \gamma_m^2 - \gamma_n^2 \right)} 
\frac{d }{d \rho} \left( \gamma_n^2 + \gamma_m^2 \right) + 
\frac{1}{2}\frac{d^2 \gamma_n^2}{d \rho^2}
\left(\frac{d \gamma_n^2}{d \rho}\right)^{-1} -
\frac{1}{2}\frac{d^2 \gamma_m^2}{d \rho^2}
\left(\frac{d \gamma_m^2}{d \rho}\right)^{-1} \right] , \\
\label{Pndanal}
& & \hskip -.8cm P_{nn}(\rho) = 
-\frac{1}{6}\frac{d^3 \gamma_n^2}{d \rho^3}
\left(\frac{d \gamma_n^2}{d \rho}\right)^{-1} +
\frac{1}{4}\left(\frac{d^2 \gamma_n^2}{d \rho^2}\right)^2
\left(\frac{d \gamma_n^2}{d \rho}\right)^{-2} \ . 
\end{eqnarray}


For correct definition of the three-body problem and solution of 
a system of HREs~(\ref{system1}), one should analyze the eigenvalues 
$\gamma_n^2(\rho )$ and matrix elements $P_{nm}(\rho )$ and 
$Q_{nm}(\rho )$, especially, near TCP (in the limit $\rho \to 0$) 
and in the asymptotic region $\rho \to \infty $. 

For $S = (-)^L$, i.~e., for odd (even) $L$ if identical particles are 
fermions (bosons), firstly, one should describe the solution 
of~(\ref{transeq1}) if $ \gamma $ tends to any integer. 
For $ \gamma \leq L + 1 $ or $ \gamma \to L + 2 n - 1 $ 
($ n \geq 1 $), the solution remains continuous, nevertheless, 
a special care is needed to take properly these limits, especially, 
in numerical calculations. 
In details, continuity at $ \gamma \leq L + 1 $ follows from Eq.~(\ref{varphi_pol}). 
In other cases, $ |\rho | $ tends to $ \infty $ for 
$ \gamma \to L + 2 n $ ($ n \geq 1 $). 
As a result, with increasing $ \rho \sign(a) $ from $ -\infty $ 
to $ \infty $, all the solutions of~(\ref{transeq1}) 
$\gamma_n^2(\rho )$ decrease monotonically from $ (L + 2n)^2 $ to 
$ (L + 2n - 2)^2 $ except for $\gamma_1^2(\rho )$, which starts from 
$ (L + 2)^2 $ and tends to $ - \infty $ as 
$\gamma^2_1(\rho) = -\rho^2 + L(L + 1) + O(\rho^{-2}) $. 
An important conclusion is that only the lowest effective channel 
potential $V_1(\rho )$ features attraction, whereas the dominant 
term $\gamma_n^2(\rho )/\rho^2$ manifests that the upper effective  
potentials $V_n(\rho )$ ($n \ge 2$) are repulsive. 

Moreover, for $S = -(-)^L$, i.~e., for even (odd) $L$ if 
identical particles are fermions (bosons), one finds that 
$\gamma_n^2(\rho )/\rho^2 \ge -1$ ($ n \geq 1 $) for any mass ratio, 
i.~e., the effective potentials in HREs exceed the two-body threshold 
$ E_\mathrm{th} = -\varepsilon_{12} = -1 $, which prohibits 
the three-body bound states. 
Hence, it is sufficient to take only $S = (-)^L$ in the study of 
the three-body bound states. 


Analysis of the wave function near TCP needs a special care as 
the channel potentials in a system of HRE $V_n(\rho )$ are singular 
for $\rho \to 0$. 
In fact, as follows from the described above properties of 
the eigenvalues and coupling terms, it is necessary to consider only 
the lowest channel potential $V_1(\rho )$.   
Its singularity is determined by the leading-order terms of 
the expansion $\gamma_1^2(\rho ) = \tilde{\gamma }^2 + q \,\rho\, 
+ O(\rho^2)$, where the notations $\tilde{\gamma } \equiv \gamma_1(0)$ 
and $q \equiv \displaystyle\left[ \frac{d\gamma_1^2(\rho )}
{d\rho} \right]_{\rho = 0}$ are introduced for brevity. 

\begin{figure}[htb]
     \begin{center}
        \subfigure[]{\label{fig_gamma}
             \includegraphics[width=0.48\textwidth]{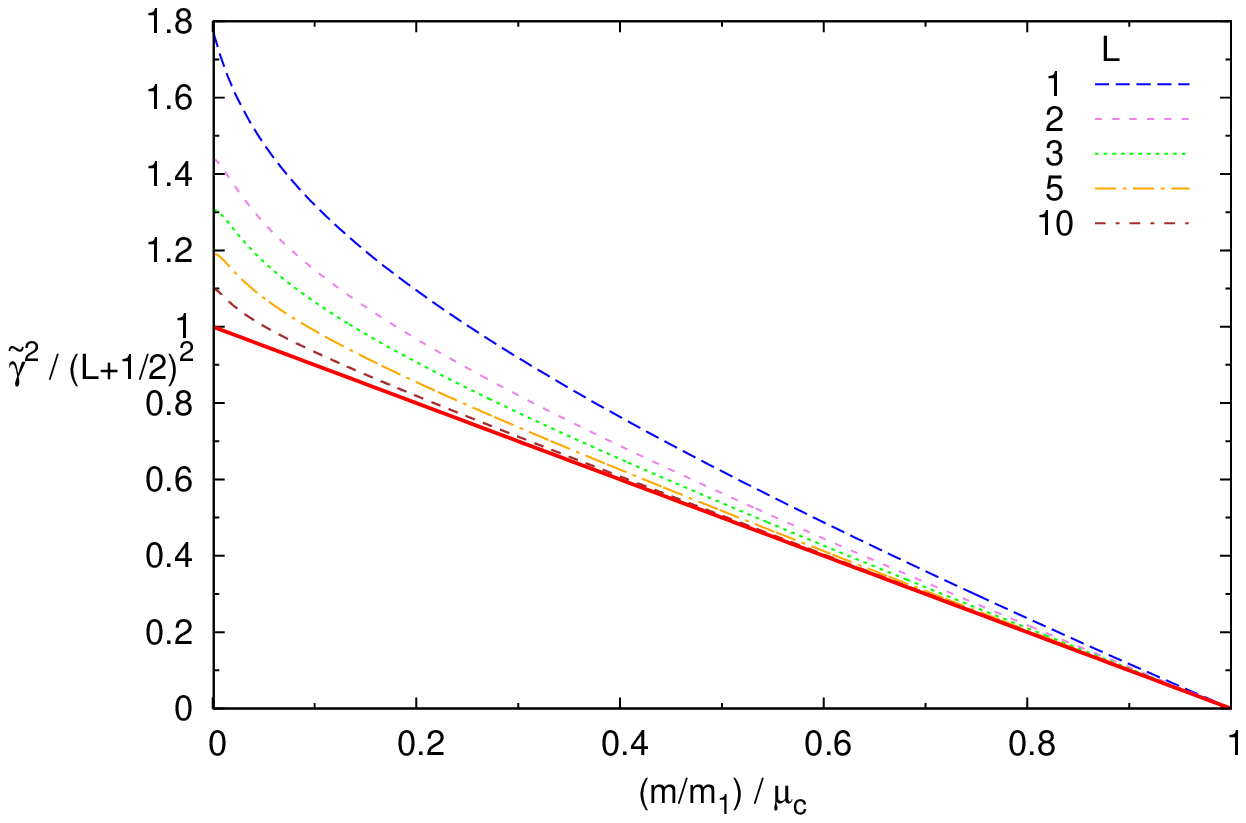}
} 
        \subfigure[]{\label{fig_q}
           \includegraphics[width=0.48\textwidth]{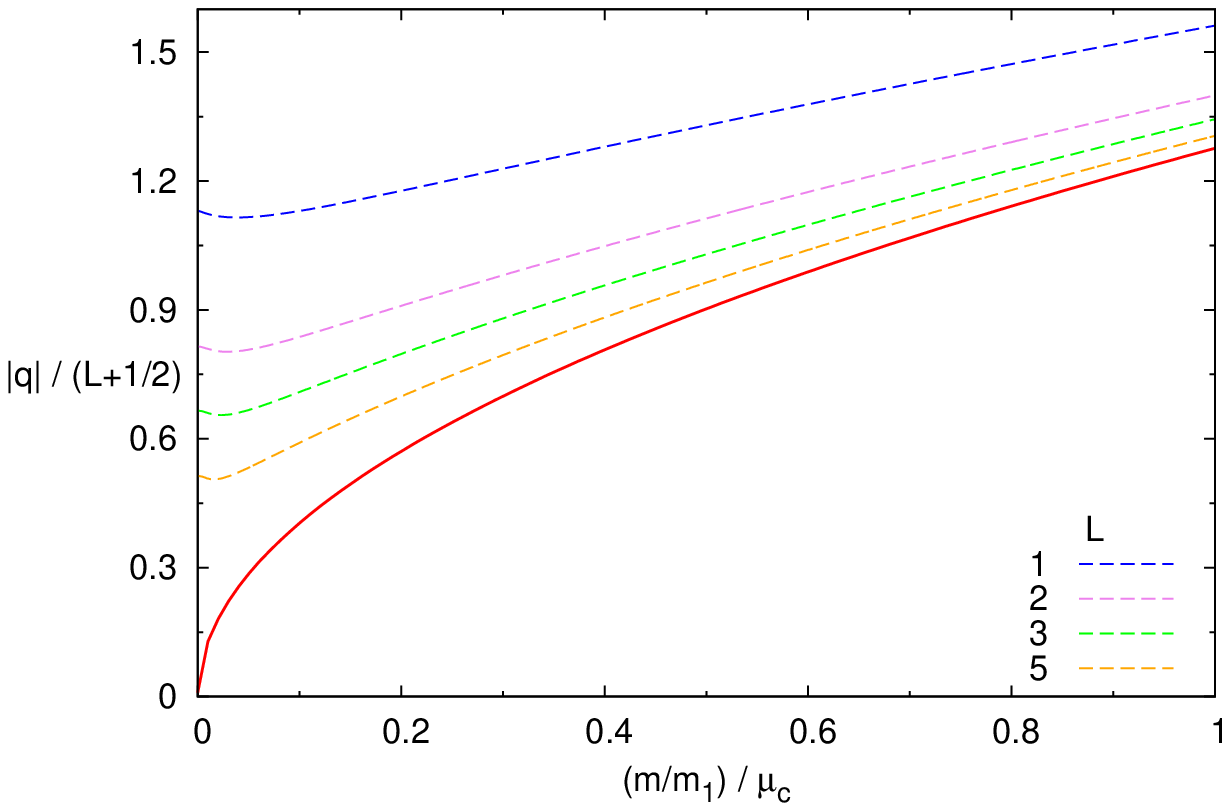}}\\ 
    \end{center}
\caption{
Mass-ratio dependencies of $ \tilde\gamma^2 $ (a) and $q$ (b)  
for two identical fermions (bosons) if $L$ is odd (even). 
In (a) the curves correspond to $ L = 1, 2, 3, 5, 10 $ (top to bottom) 
and the uppermost solid line depicts the dominant asymptotic law 
$ \tilde\gamma^2/(L + 1/2)^2 \approx 1 - \frac{m/m_1}{\mu_c} $ for 
$ L \to \infty $ as follows from Eqs.~(\ref{mucr_cer}). 
In (b) the curves correspond to $ L = 1, 2, 3, 5 $ (top to bottom)
and the uppermost solid line depicts the dominant asymptotic law 
$ |q|/(L + 1/2)^2 \approx 2/(1+u0)\sqrt{\frac{m/m_1}{\mu_c}}$ for 
$ L \to \infty $ as follows from Eqs.~(\ref{mucr_cer}) and~(\ref{qas}). 
}  
\label{fig_gammaq}
\end{figure}

If $ S = (-)^L $, one finds from~(\ref{transeq1}) that, except for 
$ L = 0 $, $ \tilde{\gamma }^2 $ monotonically decreases with 
increasing $ m/m_1 $ from $ \tilde{\gamma }^2 = (L + 1)^2 $ at 
$ m/m_1 = 0 $, passes through zero at the critical value 
$ m/m_1 = \mu_c $, and becomes negative for $ m/m_1 > \mu_c $ (pure 
imaginary $ \tilde{\gamma} $), which manifests the Efimov 
effect~\cite{Efimov73,Kartavtsev07a}. 
As for bosons in the $ L = 0 $ states, $\tilde{\gamma }^2 = 0$ is zero at 
$ m/m_1 = 0 $ and decreases with increasing $ m/m_1 $, which means 
occurrence of the Efimov effect for any finite 
masses, i.~e., $ \mu_c = 0 $. 
Along with the condition $ \tilde\gamma = 0 $ determining $ \mu_c $, 
of special importance are the values $ \tilde\gamma = 1/2 $ and 
$ \tilde\gamma = 1$ determining the critical mass-ratio values 
$ m/m_1 = \mu_e $ and $ m/m_1 = \mu_r $, respectively. 
As it will be discussed below, an additional three-body parameter is 
needed for correct formulation of the problem if $ \tilde\gamma < 1$ 
($ m/m_1 > \mu_r $) and definition of this parameter depends on whether 
$ \tilde\gamma < 1/2 $ ($m/m_1 > \mu_e$) or $ \tilde\gamma > 1/2 $  ($m/m_1 < \mu_e$). 
The dependencies $\tilde{\gamma }^2$ and $q$ on $ m/m_1 $ are shown 
in Fig.~\ref{fig_gammaq} for few lowest values of $L$ and $S = (-)^L$, 
i.~e., for fermions (bosons) if $L$ is odd (even). 
Note that $q>0$ ($q<0$) for $a>0$ ($a<0$).
The explicit equations for $ \mu_c $, $ \mu_e $, and $ \mu_r $ 
(in terms of corresponding $\omega_c$, $\omega_e$, and $\omega_r$) 
are obtained by simplification of Eq.~(\ref{transeq1}) and using Eq.~(\ref{varphi_pol}) 
\begin{subequations}
\label{omega_crit}
\begin{equation}
\label{omegac}
\Gamma^2 \left( \frac{L}{2} + 1 \right) \cos\omega_c - 
\left( \frac{\sin \omega_c }{2}\right)^L \left(\frac{1}{\sin\omega_c} 
\frac{d}{d \omega_c}\right)^L \frac{\omega_c}{\sin\omega_c} = 0 \ , 
\end{equation}
\begin{equation}
\label{omegae}
(L + 1/2)\sqrt{\sin \omega_e} \cos\omega_e -
\left( \tan \frac{\omega_e}{2}\right)^{L + 1/2} = 0 \ , 
\end{equation}
\begin{equation}
\label{omegar}
2^{L - 1}(L + 1)\Gamma^2 \left( \dfrac{L + 1}{2}\right) \cos \omega_r 
+ (\sin\omega_r )^L \left(\frac{1}{\sin\omega_r} 
\dfrac{d}{d \omega_r}\right)^L 
\left( \omega_r \cot \omega_r \right) = 0 \ . 
\end{equation}
\end{subequations} 
A list of $ \mu_r $, $ \mu_e $, and $ \mu_c $ for 
$ L = \overline{0, 10} $ is given in Table~\ref{tab1}. 
\begin{table}[htb]
\caption{The critical mass ratios $ \mu_r $, $ \mu_e $, and $ \mu_c $ 
corresponding to $\gamma = 1$, $1/2$, and $0$ for 
$ L = \overline{0, 10} $ and $ S = (-)^L $, i.~e., odd (even) 
$L$ if identical particles are fermions (bosons). } 
\label{tab1}
\begin{tabular}{ccccccc}
$L$  &  $\mu_r$&  $\mu_e$&  $\mu_c$ & $|q_r|$ & $|q_e|$ & $|q_c|$ \\
\hline
$0$ &  - &  - &  0 & - & - & \\
$1$ &  8.6185769247 &  12.313099346 &  13.606965698 &  2.0918978 &  2.2795930  &  2.3425382\\
$2$ &  32.947611782 &  37.198932993 &  38.630158395 &  3.3002049 &  3.4491653  &  3.4981899\\
$3$ &  70.070774958 &  74.510074146 &  75.994494341 &  4.5462732 &  4.6644862  &  4.7034648\\
$4$ &  119.73121698 &  124.25484012 &  125.76463572 &  5.8053122 &  5.9020644  &  5.9340526\\
$5$ &  181.86643779 &  186.43468381 &  187.95835509 &  7.0703622 &  7.1518593  &  7.1788515\\
$6$ &  256.455446   &  261.0500269  &  262.582047   &  8.346725  &  8.420731   &  8.445263\\
$7$ &  343.489658   &  348.1010286  &  349.638439   &  9.615754  &  9.679806   &  9.701066\\        
$8$ &  442.965041   &  447.5877601  &  449.128842   &  10.88657  &  10.94303   &  10.96179\\
$9$ &  554.879529   &  559.5102570  &  561.053949   &  12.15857  &  12.20906   &  12.22585\\
$10$&  679.231936   &  683.8685388  &  685.414150   &  13.43140  &  13.47707   &  13.49225\\
\end{tabular}
\end{table} 

In the opposite case $S = (-)^{L + 1}$, i.~e., for even (odd) 
$ L > 0 $ if identical particles are fermions (bosons), 
$\tilde \gamma^2 $ monotonically increases from $ (L + 1)^2 $ to 
$ (L + 2)^2 $ with increasing $0 < m/m_1 < \infty $, while for 
fermions in the $ L = 0 $ state $\tilde{\gamma }^2$ 
monotonically increases from $4$ to $16$. 

It is not surprising that the parameter $ \tilde{\gamma} $, which 
essentially determines the solution for $ \rho \to 0 $, naturally 
appears in papers \cite{Minlos11,Minlos12,Minlos14,Minlos14a,Correggi12,Correggi15,Michelangeli18,Becker18,Castin11} 
treating the problem by means of the momentum-space integral equation. 
Within the framework of this approach, the equation for $ \tilde{\gamma} $ 
(in adapted notations) reads
\begin{equation}
\label{gammaintrep}
\dfrac{(-)^L}{\pi \cos \omega } \int_{-1}^1 dx P_L(x) \int_0^{\infty} 
\dfrac{dy y^{\tilde{\gamma}}}{y^2 + 2 x y \sin \omega + 1} = 1 \ , 
\end{equation}
where $ P_L(x) $ is the Legendre polynomial. 
Evaluation of the inner integral and change of variables 
$ \sin z = - x \sin \omega $ gives 
\begin{equation}
\label{gammaintrep1}
\dfrac{2}{\sin \tilde{\gamma} \pi \sin 2\omega } 
\int_{- \omega}^{\omega} dz 
P_L \left(\dfrac{\sin z}{\sin \omega } \right) 
\sin [\tilde{\gamma} (\pi/2 - z)] = 1 \ . 
\end{equation} 
This equation is equivalent of.~(\ref{transeq1}) for $\rho = 0$.

The coupling terms of HREs are readily deduced in the asymptotic 
region $ \rho \to \infty $ from the expansion of $ \gamma_n^2(\rho) $ 
and analytical expressions~(\ref{Qanal})--(\ref{Pndanal}), 
which gives $Q_{nm}(\rho) = O(\rho^{-2})$ and $P_{nm} = O(\rho^{-4})$ 
for all $n, m$, except the lowest-channel couplings for $a > 0$, which 
are given by $P_{11}(\rho) = \frac{1}{4\rho^2} + O(\rho^{-6})$, 
$P_{1n}(\rho) = O(\rho^{-7/2})$ and $Q_{1n}(\rho) = O(\rho^{-5/2})$. 
Thus, the asymptotic form of the channel potentials is 
\begin{equation}\label{Vnaneg_inf}
V_n(\rho) = \frac{[L + 2 n - 1 - \mathrm{sgn}(a) ]^2 - 1/4}{\rho^2} + 
O(\rho^{-3}) \ , 
\end{equation} 
except of the lowest one for $a > 0$, which is given by 
\begin{equation}
\label{V1apos_inf}
V_1(\rho ) = -1 + \frac{L(L + 1)}{\rho^2} + O(\rho^{-4}) \ . 
\end{equation}

\section{Boundary conditions for generalized Coulomb problem}
\label{TCP}

For the generalized Coulomb problem the effective potential contain two 
singular terms $(\tilde\gamma^2 - 1/4)/\rho^2 + q/\rho$.  
It is necessary to analyse 
\begin{equation}
\label{eqf}
\left(\frac{d^2}{d \rho^2} - \frac{\tilde\gamma^2 - 1/4}{\rho^2} - 
\frac{q}{\rho} + E \right)f(\rho) = 0
\end{equation}
at small $\rho$. 
There are two solutions at $\rho \to 0$, which leading order terms are  
$\sim \rho^{1/2 \pm \tilde\gamma}$. 

For $\tilde\gamma^2 \ge 1$, $f(\rho)\sim \rho^{1/2 + \tilde\gamma}$ is 
the only square-integrable solution for $\rho \to 0$ (the appropriate 
boundary condition $f(\rho) \to 0$). 
In contrast, for $\tilde\gamma^2 < 1$ both solutions $\sim \rho^{1/2 \pm \tilde\gamma}$ at $\rho \to 0$ are square-integrable. 
Therefore, for unambiguous formulation of the three-body problem 
near $\rho \to 0$ it is necessary to fix the linear combination of these 
solutions in $f(\rho)$, which requires the additional three-body 
parameter. 
If $\tilde\gamma^2 < 0$ there is an Efimov situation, namely, both 
square-integrable solutions at $\rho \to 0$ are oscillating. 
And it is already known that the additional three-body regularizational 
parameter (called the Efimov parameter) is needed to fix the wave function 
in the TCP that results in energy spectrum exponentially depending 
on the level's number 
$E = - e ^{-\frac{2 \pi} {|\tilde \gamma|} n}$~\cite{Case50}.  
Furthermore, one should consider the remaining case 
$0 \leq \tilde\gamma^2 < 1$. 
As follow from Eq.~(\ref{eqf}), on the interval $1/2 < \tilde\gamma < 1 $ 
($\mu_r > m/m_1 > \mu_e$) one has to take into account also the next to 
leading order term in the second square-integrable solution 
$\sim \rho^{1/2 - \tilde\gamma}$, namely, 
$q \rho^{3/2 - \tilde\gamma} /(1 - 2 \tilde\gamma )$, because it is of 
the same order as the first square-integrable solution 
$\sim \rho^{1/2 + \tilde\gamma}$. 
As a result, denoting the additional three-body parameter via $b$, 
the three-body boundary condition for the channel function $f(\rho)$ 
read as
\begin{equation}
\label{as_gam121}
\displaystyle f(\rho)
\xrightarrow[\rho \to 0]{}
\rho^{1/2 + \tilde\gamma} - \sign (b) |b|^{2 \tilde\gamma} 
\rho^{1/2 - \tilde\gamma} 
\left[1 + q \rho /(1 - 2 \tilde\gamma )\right] \  
\end{equation}
for all $\tilde\gamma$ except for $\tilde\gamma = 1/2$. 
The last term in the square brackets ($\sim q \rho$) is necessary only 
for $1/2 < \tilde\gamma < 1 $ and can be omitted for 
$0 < \tilde\gamma < 1/2 $. 
In the limit $\tilde\gamma \to 0$, the boundary condition~(\ref{as_gam121}) 
takes a simple form 
\begin{equation}
\label{as_gam0}
f(\rho) \xrightarrow[\rho \to 0]{} \rho^{1/2} \log (\rho /b) \ , 
\end{equation}
where only $b > 0$ is allowed. 
In the specific case $\tilde\gamma = 1/2$ there are two square-integrable 
solutions at $\rho \to 0$, namely, $\rho$ and $1 + q \rho \log \rho$. 
The boundary condition reads  
\begin{equation}
\label{as_gam12}
f(\rho) \xrightarrow[\rho \to 0]{} \rho - b (1 + q \rho \log \rho) \ . 
\end{equation}

It is suitable to write the three-body boundary conditions in 
the alternative form,~viz., in terms of the derivative of the function 
$f(\rho)$. 
The boundary condition for $0 \leq \tilde\gamma < 1$  $\&$ 
$\tilde\gamma \neq 1/2$ reads 
\begin{equation} 
\label{bc_gam121}
\displaystyle\lim_{\rho \to 0} \left( \rho^{1 - 2\tilde\gamma } 
\frac{d}{d\rho } + \sign (b) \frac{2\tilde\gamma}{|b|^{2\tilde\gamma}}\right) 
\frac{\rho^{\tilde\gamma - 1/2}} {1 - 2\tilde\gamma + q\rho}  
f(\rho) = 0 \ ,
\end{equation} 
which is equivalent to Eq.~(\ref{as_gam121}). 
In the limit $\tilde\gamma \to 0$ the boundary condition, which is equivalent 
to Eq.~(\ref{as_gam0}), takes the form 
\begin{equation} 
\label{bc_gam0}
\displaystyle \lim_{\rho \to 0} \left( \rho \frac{d}{d\rho } - 
\frac{1}{\log (\rho /b)}\right) \rho^{-1/2}f(\rho) = 0\ ,
\end{equation}
where only $b > 0$ is allowed. 
In the specific case of $\tilde\gamma = 1/2$ the boundary condition 
\begin{equation} 
\label{bc_gam12}
\displaystyle\lim_{\rho \to 0} \left( \frac{d}{d\rho} + \frac{1}{b}\right) 
\frac{f(\rho)}{1 + q\rho\log \rho} = 0\ 
\end{equation} 
is equivalent to Eq.~(\ref{as_gam12}). 
Notice that the boundary condition for $\tilde\gamma = 0$ determined by 
Eq.~(\ref{as_gam121}) or Eq.~(\ref{bc_gam0}) is similar to that for 
the $2D$ zero-range model~\cite{Kartavtsev06}, whereas for 
$\tilde\gamma = 1/2$ the boundary condition of the form~(\ref{as_gam12}) 
or~(\ref{bc_gam12}) is similar to that for a sum of the zero-range and 
Coulomb potentials as in~\cite{Yakovlev13}.
Usage $\ln (|q|\rho)$ instead of $\ln \rho$ in~(\ref{as_gam12}) for 
$\tilde\gamma = 1/2$ simply means the redefinition of the parameter $b$ 
on $\tilde {b} = b/(1 +b q \ln |q|)$, then it coincides
with~\cite{Albeverio83}. 

The usage of the boundary conditions~(\ref{as_gam121})-(\ref{as_gam12}), 
or~(\ref{bc_gam121})-(\ref{bc_gam12}) with arbitrary three-body parameter 
$b$ (with dimension of length) determine the general zero-range three-body 
potential. 
The relation of the general approach for zero-range three-body potential 
with particular examples of the shrinking three-body potentials is given 
in Appendix~\ref{Appendix_bc}. 

\section{Self-ajoint Hamiltonian}

Singular terms in the HREs~(\ref{system1}) for $\rho \to 0$ shows that 
one should apply the analysis of Sec.~\ref{TCP} to formulate of 
the three-body boundary conditions. 
The wave function $\Psi$ near the TCP ($\rho \to 0$) is basically 
determined by the most singular terms in the effective potential 
in the first channel $V_1(\rho)$~(\ref{Vef}),~i.~e., 
$ (\tilde\gamma^2 - 1/4) / \rho^2 + q / \rho$ and the corresponding 
channel function $f_1(\rho)$. 
In the following the general analyse of Sec.~\ref{TCP} will be applied  
to the first channel $f(\rho)\equiv f_1(\rho)$. 
Therefore, using the analysis of behavior of $\tilde \gamma$ and $q$ as 
functions of mass ratios and using the way of regularization of the wave 
functions of Sec.~\ref{TCP} in dependence on $\tilde \gamma$, 
one comes to the conclusions.
First, one finds that for the mass-ratios $m/m_1 \le \mu_r$ and 
$S=(-1)^L$ (odd or even  $L$ if identical particles are fermions or bosons, 
respectively) and for any mass ratio and $S=(-1)^{L+1}$ (even or odd $L$ if 
identical particles are fermions or bosons, respectively) the Hamiltonian is 
self-ajoint, due to $\tilde \gamma > 1$.
One can use the zero boundary condition in TCP. 
Second, notice that the Efimov situation corresponds to $m/m_1 \ge \mu_c$ 
and $S=(-1)^L$ due to $\tilde \gamma \le 0$. 
The way of regularisation in the TCP to make the Hamiltonian self-ajoint is 
well investigated in the literature. 
Third, the most interesting case $0 \le \tilde \gamma < 1$ corresponds 
to the mass-ratios $\mu_r < m/m_1 \le \mu_c$ and $S=(-1)^L$ (odd or even 
$L$ if identical particles are fermions or bosons, respectively). 
To make the self-ajoint Hamiltonian one need to use the corresponding 
boundary condition in the TCP of the form~(\ref{as_gam121})--(\ref{as_gam12}), 
or~(\ref{bc_gam121})--(\ref{bc_gam12}) for the channel function in 
the first channel. 
Only this case will considered below. 

Remark that the boundary condition in~\cite{Michelangeli18} coincides with 
Eq.~(\ref{as_gam121}) or~(\ref{bc_gam121}) if one omit the term $\sim q$, 
that means that the boundary condition in~\cite{Michelangeli18} can be 
used only for $0 \ge \tilde\gamma \ge 1/2$. 
The three-body parameter $b$ ($R_t$ in~\cite{Gao15}) and the three-body 
boundary condition was introduced also in paper~\cite{Gao15} devoted to 
calculation of the third virial coefficient, where only positive value of 
$b$ is taken into account and the term $\sim \rho^{-1}$ is not considered. 
For $0 \leq \tilde\gamma < 1/2$ this term is of the principal importance. 
Generally, the three-body parameter $b$ and the boundary condition can 
depend not only on $L$, but on it's projection $M$. 
Nevertheless, in real situation it doesn't seem probable. 

It is of interest to write boundary conditions for the total wave function 
$\Psi$. 
Namely, as in~\cite{Kartavtsev16}, the required expressions can be 
written as 
\begin{equation}
\Psi\sim(\rho^{\tilde\gamma-2}\mp |b|^{2 \tilde\gamma}\rho^{-\tilde\gamma-2}) 
\Phi_1(0, \Omega)
\end{equation} 
or  
\begin{equation} 
\label{psibc_gam120}
\displaystyle\lim_{\rho \to 0} \left( \rho^{1 - 2\tilde\gamma } \frac{d}{d\rho } 
\pm \frac{2\tilde\gamma}{|b|^{2\tilde\gamma}}\right) \rho^{2 + \tilde\gamma}\Psi = 0 \ , 
\end{equation}
if $0 < \tilde\gamma < 1/2$ that equivalent to~(\ref{as_gam121}),~(\ref{bc_gam121}). 
On the other hand, the boundary condition for $1/2 < \tilde\gamma < 1$ becomes 
cumbersome due to necessity to keep in the expansion of $\Psi$ for 
$\rho \to 0$ also the term $\sim \rho^{-\tilde\gamma-1}$, 
which includes an additional function of hyper-angles.

\section{Bound-state energies}
\subsection{Infinite two-body scattering length}

In the limit $|a| \to \infty$, $\gamma_n^2(\rho)$ in~(\ref{transeq1}) 
do not dependent on $\rho$ and all the terms $Q_{nm}(\rho)$ and 
$P_{nm}(\rho)$ vanish.
Therefore, HREs~(\ref{system1}) decouple and the three-body bound-state 
energies is a solution of one HRE, in which 
$\gamma^2(\rho) \equiv \tilde\gamma^2$. 
For $b > 0$, there is one bound state whose energy is
$ E = -4b^{-2}\left[-\Gamma(\tilde\gamma )/\Gamma(-\tilde\gamma )
\right]^{1/\tilde\gamma }$ 
and eigenfunction is $f(\rho) = \rho^{1/2} K_{\tilde\gamma }(\sqrt{-E} \rho)$,
where $K_\nu(x)$ is the modified Bessel function. 
If $b \to \infty$, the bound state goes to the threshold $E\to 0$ and 
turning into the virtual state.
Then, for $b < 0$ it's energy is given by the above expression. 
Also, the above expressions for $E$ and $f(\rho)$ describe the properties 
of the bound deep state, which exists for $|a| \gg b$. 

\subsection{Simple model}
As a preliminary consideration, it is worthwhile to give qualitative 
description of the energy spectrum as function of $b$ and $m/m_1$ within 
the framework of the simple model. 
The model is equivalent to the generalised Coulomb problem incorporating 
the zero-range interaction and is based on splitting of the Hamiltonian 
into the singular part $(\gamma^2 - 1/4)/\rho^2 + q/\rho$ as 
$\rho \to 0$ and the remaining one, which is simply taken as a constant  
$\epsilon(\gamma)$ smoothly dependent on $m/m_1$. 
Retaining one equation containing the most singular terms from 
the system~(\ref{system1}), one comes to the equation 
\begin{equation}
\displaystyle \left(\frac{d^2}{d \rho^2} - \frac{\gamma^2 - 1/4}{\rho^2} - 
\frac{q(\gamma)}{\rho} + E - \epsilon (\gamma) \right)f(\rho) = 0 
\end{equation} 
complimented by one of the boundary conditions~(\ref{as_gam121}), 
(\ref{as_gam0}), and (\ref{as_gam12}).
Similar to~\cite{Kartavtsev16}, the solution of generalised Coulomb 
problem leads to the eigenenergy equations 
\begin{subequations}
\label{traneq_simple_model}
\begin{equation}\label{traneq_simple_model_a}
\left(2\kappa |b|\right)^{2\gamma } = 
\displaystyle \mp \frac{\Gamma (2 \gamma ) 
\Gamma \left(1/2 - \gamma + q/(2 \kappa) \right)} 
{\Gamma (-2 \gamma ) \Gamma \left(1/2 + \gamma + q/(2 \kappa) \right)} 
\quad 0 \leq \gamma < 1, \gamma \ne 1/2,
\end{equation}
\begin{equation}\label{traneq_simple_model_b}
\ln (2 \kappa b) + \psi \left(\frac{1}{2} + \frac{q}{2\kappa} \right) + 
2\gamma_C  = 0, \qquad\qquad\qquad  \gamma = 0, \quad b \geq 0,
\end{equation}
\begin{equation}\label{traneq_simple_model_c}
\frac{1}{q}\left( \frac{1}{ b} - \kappa\right) - 
\ln \left( \frac{|q|}{2 \kappa}\right)+ 
\psi \left(1 + \frac{q}{2 \kappa}\right) + 
2 \gamma_{C} - 1 = 0,\quad \gamma=1/2, 
\end{equation}
\end{subequations}
where $\kappa = \sqrt{\epsilon(\gamma) - E}$, $\psi(x)$ is the digamma 
function and $\gamma_{C} \approx 0.5772$ is the Euler--Mascheroni 
constant.
Eq.~(\ref{traneq_simple_model_b}) can be obtained from 
Eq.~(\ref{traneq_simple_model_a}) by taking the limit $\gamma\to 0$ for 
any $b \geq 0$. 
Recall that the parameter $b$ in Eq.~(\ref{traneq_simple_model_c}) is 
defined in Eq.~(\ref{as_gam12}) differently. 

As follows from Eqs.~(\ref{traneq_simple_model}), all the bound-state 
energies monotonically increase with increasing $b$; moreover, one bound 
state arises at $-\infty$ if $b$ passes through zero. 
Particularly, in two limits $b = 0$ and $b \to \infty$ one obtains the 
Coulomb spectrum for energies
\begin{equation}
\label{Enb0inf}
E_n^{(b = 0, \infty)}(\gamma)  = 
-\frac{q^2(\gamma)}{ [2 (n + s \gamma ) + 1]^2} + 
\epsilon(\gamma) \ , 
\end{equation}
where $s = \pm 1$ corresponds to $b = 0$ ($s = +1$) and $b \to \infty$ 
($s = -1)$.
In each case the index $n \geq 0$ enumerating energy levels is limited 
either by the condition $n > - s \gamma - 1/2 $ if $a > 0$ ($q < 0$) or 
$n < - s \gamma  - 1/2$ if $a < 0$ ($q > 0$). 
The maximum value of $n$ is restricted by 
$E_n^{(b = 0, \infty)}(\gamma) < -1$ if $a > 0$ or 
$E_n^{(b = 0, \infty)}(\gamma) < 0$ if $a < 0$. 
The Eq.~(\ref{Enb0inf}) is valid for any mass ratio including 
the exceptional value $m/m_1 \to \mu_e$ ($\gamma \to 1/2$). 

The specific feature of the Coulomb spectrum~(\ref{Enb0inf}) is a 
degeneracy of energy levels for integer and half-integer value of  
$\gamma$,~i.e., at $m/m_1 \to \mu_c$ ($\gamma \to 0$), $m/m_1 \to \mu_e$ 
($\gamma \to 1/2$), and $m/m_1 \to \mu_r$ ($\gamma \to 1$). 
In the case $a > 0$, $E_{n}^{(0)}(0)=-q^2(0)/(2 n + 3)^2 + \epsilon(0)$ 
($n \geq 0$) coincides with $E_{n}^{(\infty)}(0)$ for $m/m_1 \to \mu_c$ 
($\gamma \to 0$), $E_{n}^{(0)}(1/2)=-q^2(1/2)/(2 n + 3)^2 + \epsilon(1/2)$ 
($n \geq 0$) coincides with $E_{n+1}^{(\infty)}(1/2)$ for 
$m/m_1 \to \mu_e$ ($\gamma \to 1/2$), 
$E_{n}^{(0)}(1)=-q^2(1)/(2 n + 3)^2 + \epsilon(1)$ ($n \geq 0$) coincides 
with $E_{n+2}^{(\infty)}(1)$ for $m/m_1 \to \mu_r$ ($\gamma \to 1$). 
The ground state $E_{0}^{(\infty)}(\gamma)$ tends to $-\infty$ for 
$m/m_1 \to \mu_e$ ($\gamma \to 1/2$) and disappears for $m/m_1 \leq \mu_e$ 
($\gamma \geq 1/2$). 
For $m/m_1 < \mu_e$ ($\gamma > 1/2$) $E_{1}^{(\infty)}(\gamma)$  becomes 
a ground state and for $m/m_1 \to \mu_r$ ($\gamma \to 1$) tends to a 
finite value $-q^2(1) + \epsilon(1)$, which is not degenerate with any 
$E_{n}^{(0)}(1)$.
In the case $a < 0$, there is only $E_{0}^{(\infty)}(\gamma)$ in 
the interval $\mu_r < m/m_1 < \mu_e$, which tends to 
$-q^2 (1)+ \epsilon(1)$ for $m/m_1 \to \mu_r$ ($\gamma \to 1$) and to 
$-\infty$ for $m/m_1 \to \mu_e$ ($\gamma \to 1/2$). 
One should note that both $E_{0}^{(\infty)}(1)$ for $a < 0$ and 
$E_{1}^{(\infty)}(1)$ for $a > 0$ coincides in the limit $m/m_1 \to \mu_r$ 
($\gamma \to 1$).

Furthermore, in the case $a > 0$, the energy of the $n^{th}$ level 
($n \geq 0$) for any $b \leq 0$ converges to 
$E_{n}^{(0)}(0) = -q^2(0)/(2 n + 3)^2 + \epsilon(0) =E_{n}^{(\infty)}(0)$ 
in the limit $m/m_1 \to \mu_c$ ($\gamma \to 0$). 
For $m/m_1 = \mu_c$ ($\gamma = 0$) the ground-state energy increases 
from $-\infty$ to $E_{0}^{(0)}(0)$ with increasing $b$ from zero to 
infinity, while the $n^{th}$ level increases from $E_{n-1}^{(0)}(0)$ to 
$E_{n}^{(0)}(0)$, and the upper level disappears at the threshold for 
some finite value $b>0$. 
If the mass ratio tends to the next specific value $m/m_1 \to \mu_e$ 
($\gamma \to 1/2$), for any $b$ all the energies converge to 
$E_{n}^{(0)}(1/2) = -q^2(1/2)/(2 n + 3)^2 + \epsilon(1/2) = 
E_{n+1}^{(\infty)}(1/2)$ ($n \geq  0$), and additionally the ground-state 
energy for $m/m_1 > \mu_e$  tends to $-\infty$ in the same limit. 
If the mass ratio tends to $m/m_1 \to \mu_r$ ($\gamma \to 1$), for any $b$ 
the energies converge to either 
$E_{1}^{(\infty)}(1)=-q^2(1) + \epsilon(1)$ or 
$E_{n}^{(0)}(1) =-q^2(1)/(2 n + 3)^2 + \epsilon(1) = E_{n+2}^{(\infty)}
(1)$ ($n \geq 0$). 
In the case $a < 0$, for $m/m_1 \to \mu_r$ ($\gamma \to 1$) the energies 
converge to $E_{0}^{(\infty)}$ for any $b$. 
The descriptions of the spectrum by means of the simple model are in 
agreement  with numerical calculation as can be seen in Fig.~\ref{fig_en}. 
 
A comparison of the ground and excited states energies for 
$b = 0$~\cite{Kartavtsev07a} with Eq.~(\ref{Enb0inf}) shows that 
reasonable agreement could be obtained for $\epsilon$ about 
$-0.4 \div -0.6$ for $1 \le L \le 5$. 
Using the above estimate for the constant $\epsilon$, 
one finds that for $a > 0$ there are about $L + 1$ levels below 
the two-body threshold ($E \le -1$) if $b = 0$ and about $L + 2$ levels 
if $b \to \infty$, while for $a < 0$ there is one level below 
the three-body threshold ($E \le 0$) if $b \to \infty$ (see 
Fig.~\ref{fig_en}). 

\subsection{Numerical results for $L = 1 \div 5$}

\begin{figure*}[htb] 
\hspace{-.5cm}
\includegraphics[width=0.45\textwidth]{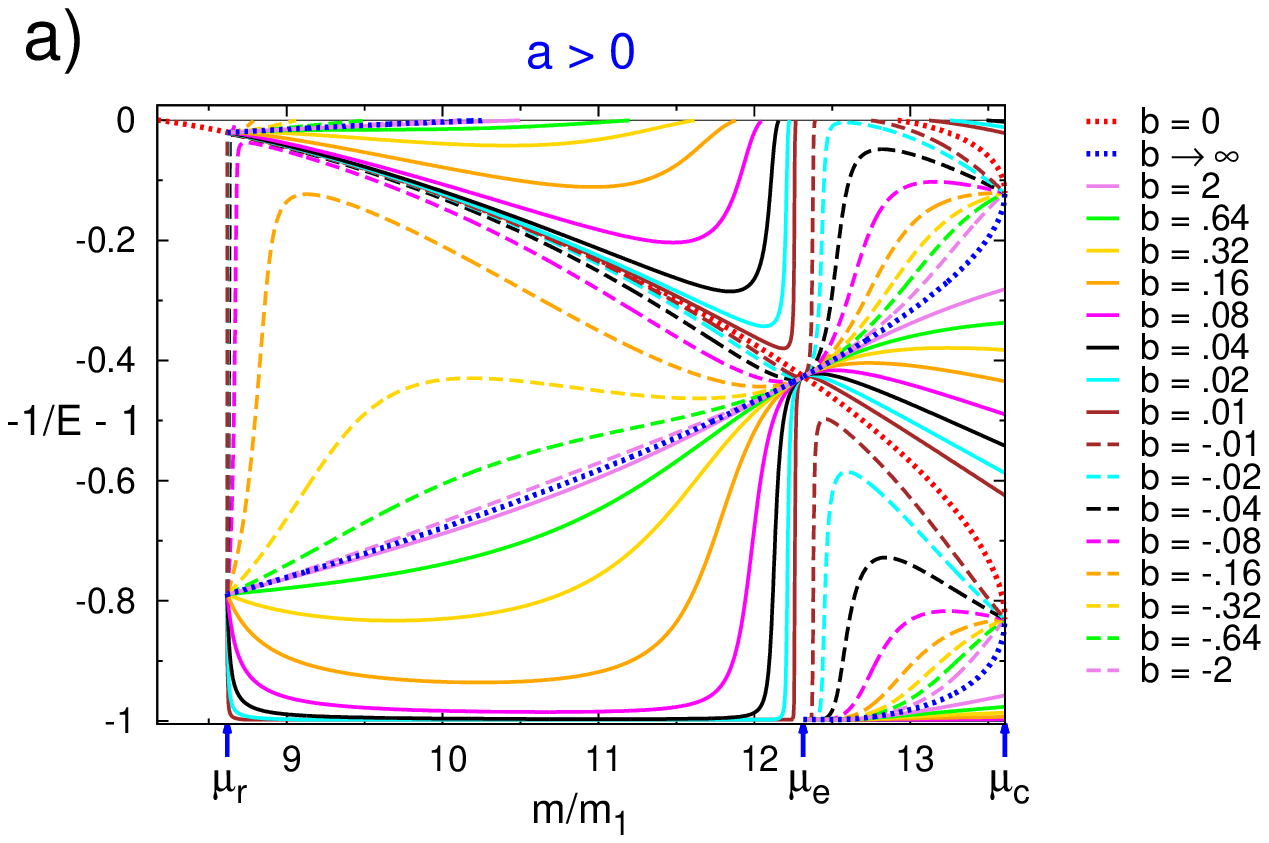} 
\includegraphics[width=0.45\textwidth]{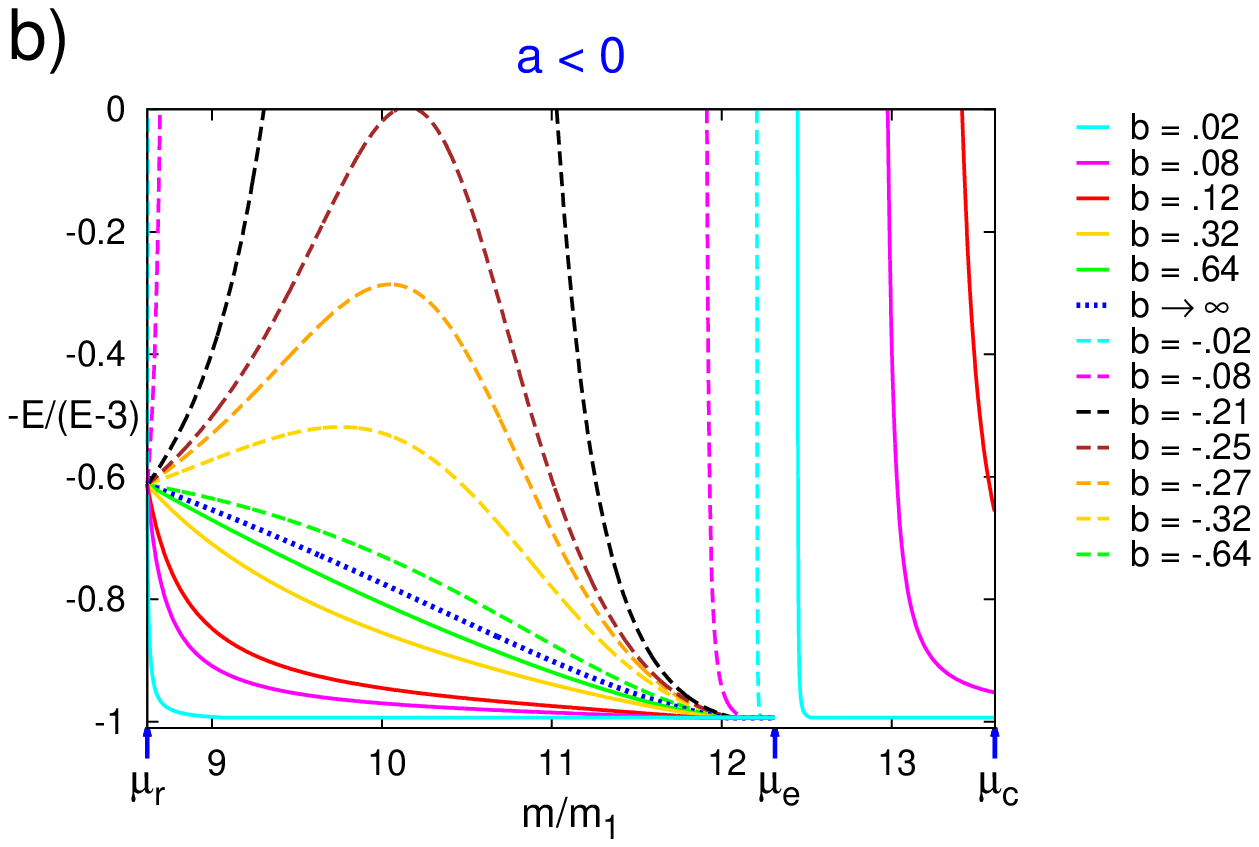}\\
\includegraphics[width=0.45\textwidth]{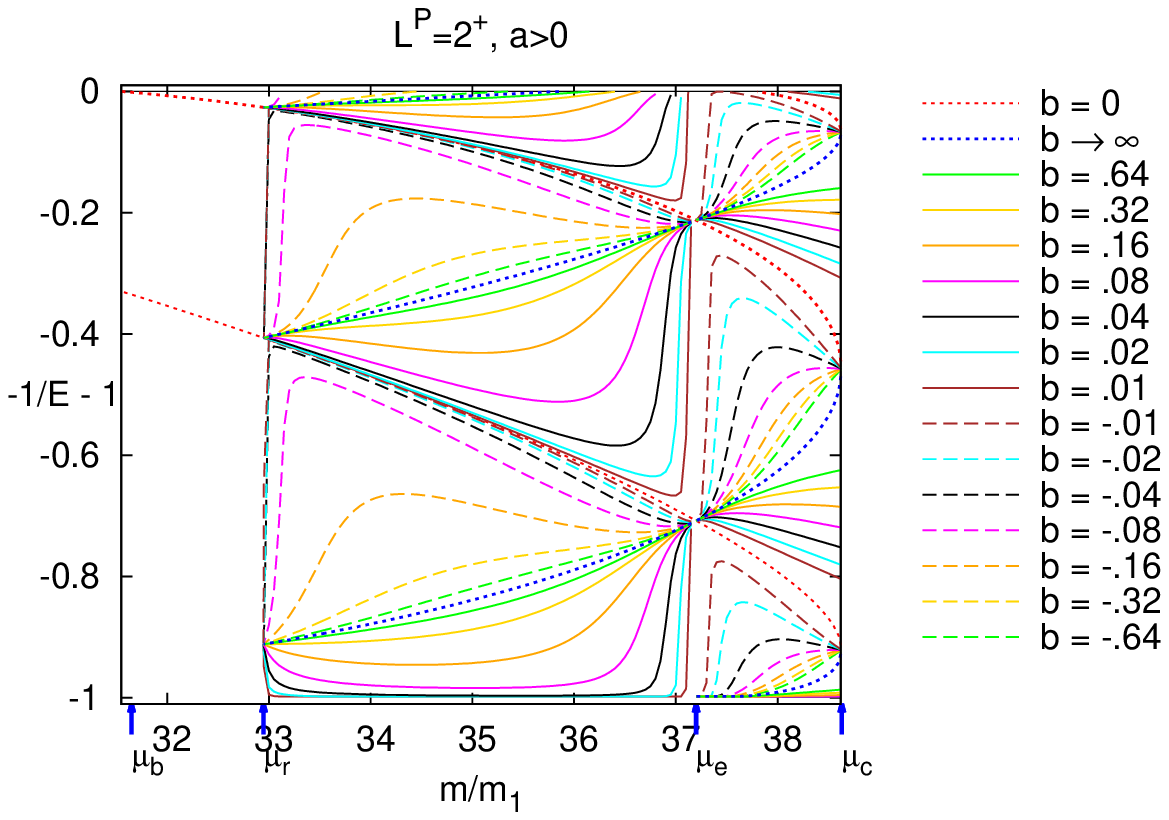} \hspace{-.7cm} 
\includegraphics[width=0.45\textwidth]{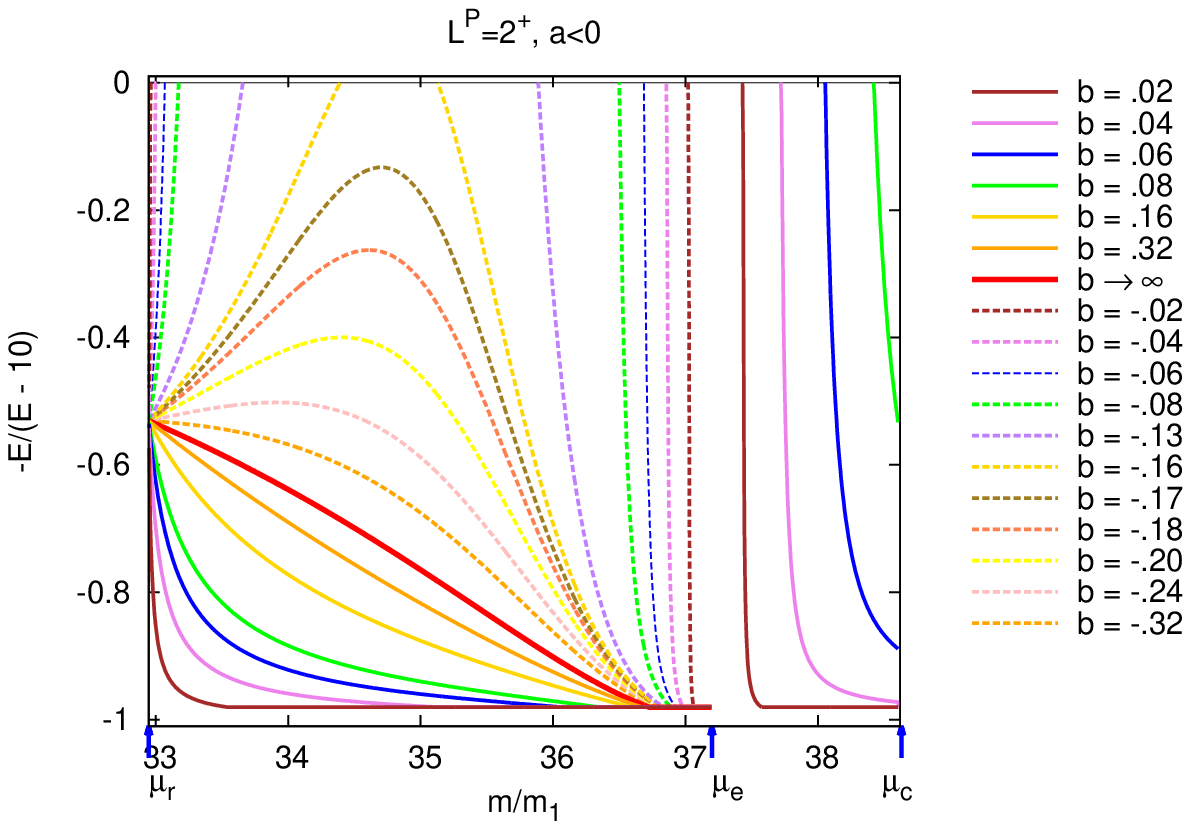}
\includegraphics[width=0.45\textwidth]{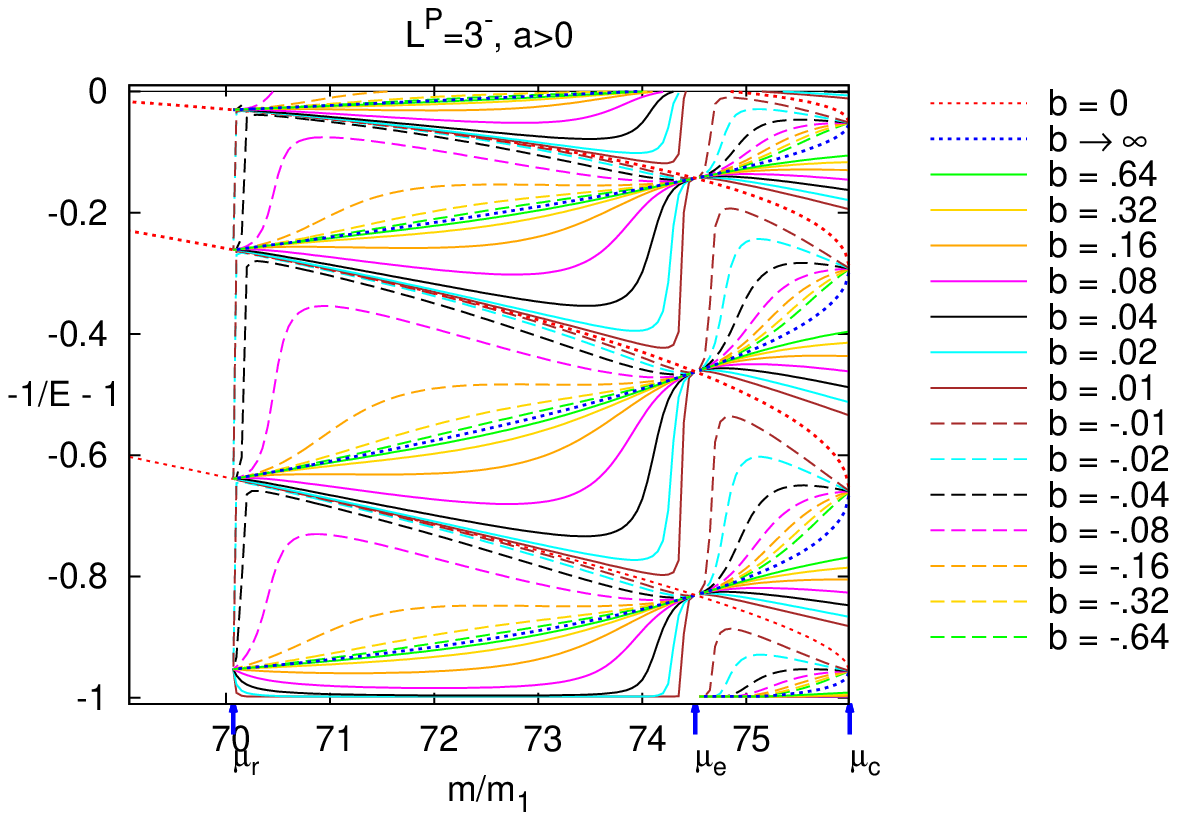} \hspace{-.7cm} 
\includegraphics[width=0.45\textwidth]{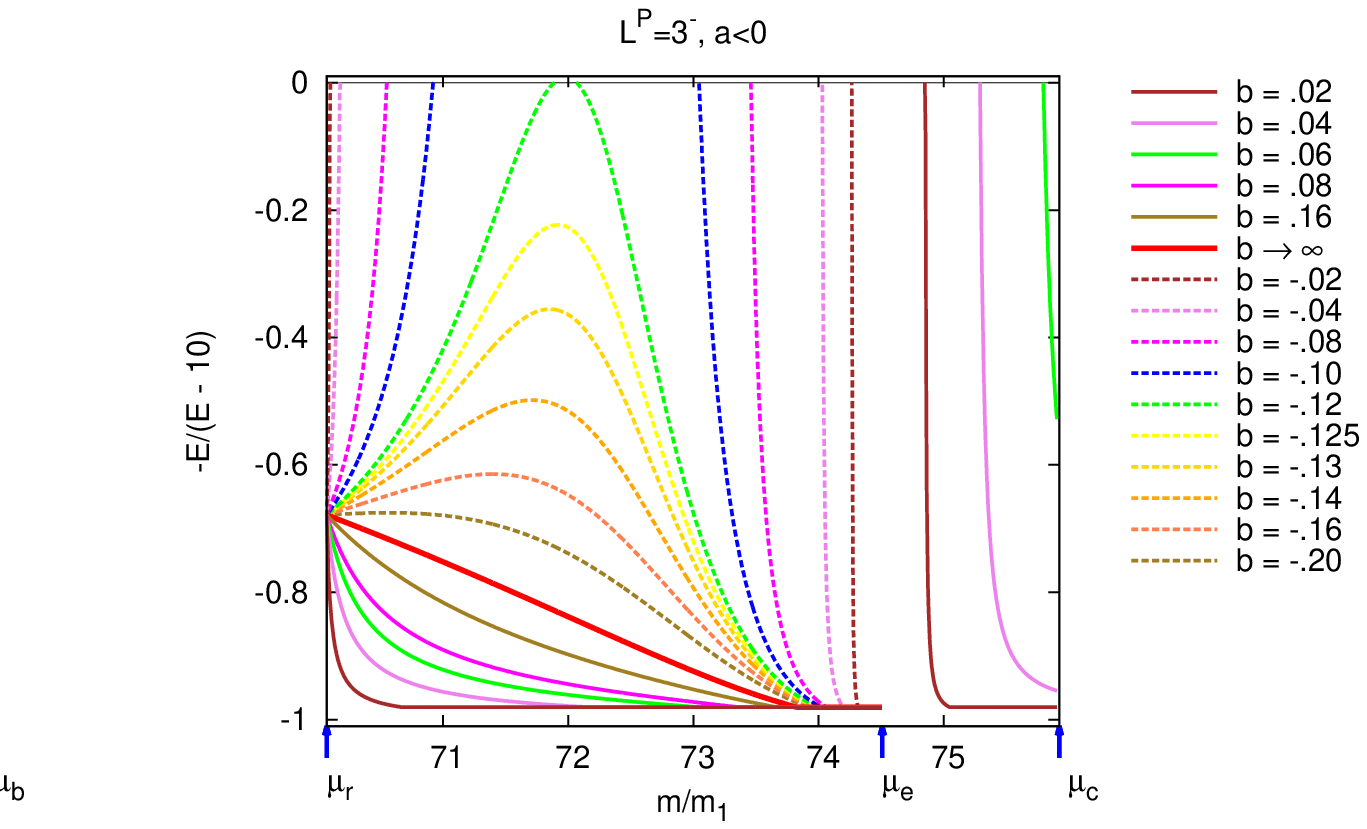}
{\caption{\scriptsize Bound-state energies $E$ for three-body system 
with two identical fermions (bosons) in $L^p = 1^-,\,  3^-$ 
(for $L^p = 2^+$) states as a function of $m/m_1$ and $b$. 
The energies for the two-body scattering length $a > 0$ and $a < 0$ are 
presented in panel a and panel b, respectively, and the energy axis scaled 
to map $-\infty < E < -1$ (panel a) and $-\infty < E < 0$ (panel b) to 
the interval $(-1, 0)$. 
Values $\mu_r$, $\mu_e$ and $\mu_c$ correspond to $\gamma = 1$, $1/2$ and 
$0$.
} 
\label{fig_en}}
\hspace{-.5cm}
\end{figure*}

The mass-ratio dependence of the three-body energies for 
angular momentum $L \le 5$  ($L^P = 1^-,\, 3^-,\, 5^-$ or 
$L^P = 2^+,\, 4^+$ if two identical particles are fermions or bosons, 
respectively) is determined on the mass ratio interval  
$\mu_r(L^P) < m/m_1 \le \mu_c(L^P)$ ($1 > \gamma \ge 0$) by solving 
a system of HREs~(\ref{system1}) complemented by the special boundary 
conditions~(\ref{as_gam121}) or~(\ref{as_gam0}) or~(\ref{as_gam12}) 
in the TCP and the zero asymptotic boundary condition, 
$f_n(\rho)\to 0$ as $\rho\to\infty$. 
Solution of up to eight HREs provides five - six digits in the calculated 
energy. 
The results of the calculations are shown in Fig.~\ref{fig_en} 
for $L^P = 1^-,\ 2^+,\ 3^-$  in the cases of positive and negative 
two-body scattering length $a$. 

For positive two-body scattering length it turns out that the number of 
the bound states increases with increasing $L$, but, qualitatively, energy 
dependence on $m/m_1$, $b$ is similar for different $L$. 
Namely, one obtain the spider-like plot with monotonic increasing of 
the bound-state energies with increasing $b$ at fixed $m/m_1$.  
In fact, the bound-state energies dependence for positive and negative 
values of $b$ are separated of each other by the limiting mass-ratio 
dependences for $b = 0$ and $b = \infty$, and by the critical value 
of mass ratio $\mu_{e}(L^P)$. 
Moreover, limiting dependences of the three-body bound-state energies 
monotonically decrease for $b = 0$ and monotonically increase for 
$b = \infty$ with increasing mass ratio as illustrated in 
Fig.~\ref{fig_en}. 
When $m/m_1$ tends to either of specific values $\mu_{r}(L^P)$, 
$\mu_{e}(L^P)$, and $\mu_{c}(L^P)$, 
the three-body energies for $b=0$ coincide with those for $b \to \infty$. 
Besides, for $b \to \infty$ there is the ground state, which energy tends 
to the finite value as $m/m_1 \to \mu_{r}(L^P)$ and to minus infinity as 
$m/m_1 \to \mu_{e}(L^P)$. 
The three-body energies for $b\neq 0,\,b\neq\infty$ tends to those 
for $b \to \infty$ in the limit of mass ratios $\mu_{r}(L^P)$, 
$\mu_{e}(L^P)$, and $\mu_{c}(L^P)$ except the positive values of $b$ 
in the limit $m/m_1\to \mu_{c}(L^P)$. 
The calculated three-body energies in three limits $\mu_{r}(L^P)$, 
$\mu_{e}(L^P)$, and $\mu_{c}(L^P)$ are presented 
in Tab.~\ref{tab2}. 

For negative two-body scattering length, the energy dependence on $m/m_1$, 
$b$ is similar for different $L^P$ sectors: the only bound state exists 
for any positive value of $b$ and some negative values of $b$ on 
the mass-ratio interval  $\mu_r(L^P) < m/m_1 \le \mu_e(L^P)$, whereas 
on the interval $\mu_e(L^P) < m/m_1 \le \mu_c(L^P)$ the bound state exists 
only for small enough positive values of $b$. 
Three-body bound-state energies for limiting value of the mass ratio 
$\mu_{r}(L^P)$ are presented  as underlined numbers in 
the Tab.~\ref{tab2}. 
They considers with the same numbers for positive $a$.

\begin{figure*}[htb]
\hspace{-.5cm}
\includegraphics[width=.52\textwidth]{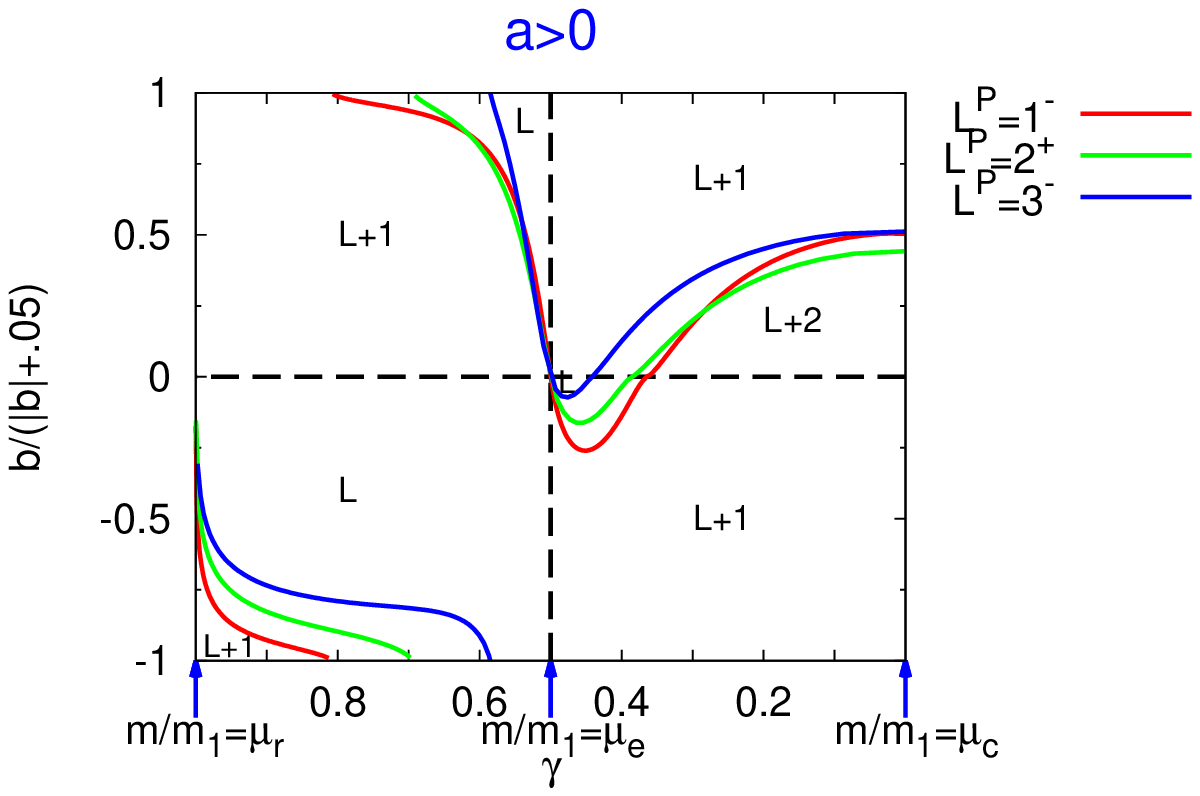} \hspace{-.7cm} 
\includegraphics[width=.52\textwidth]{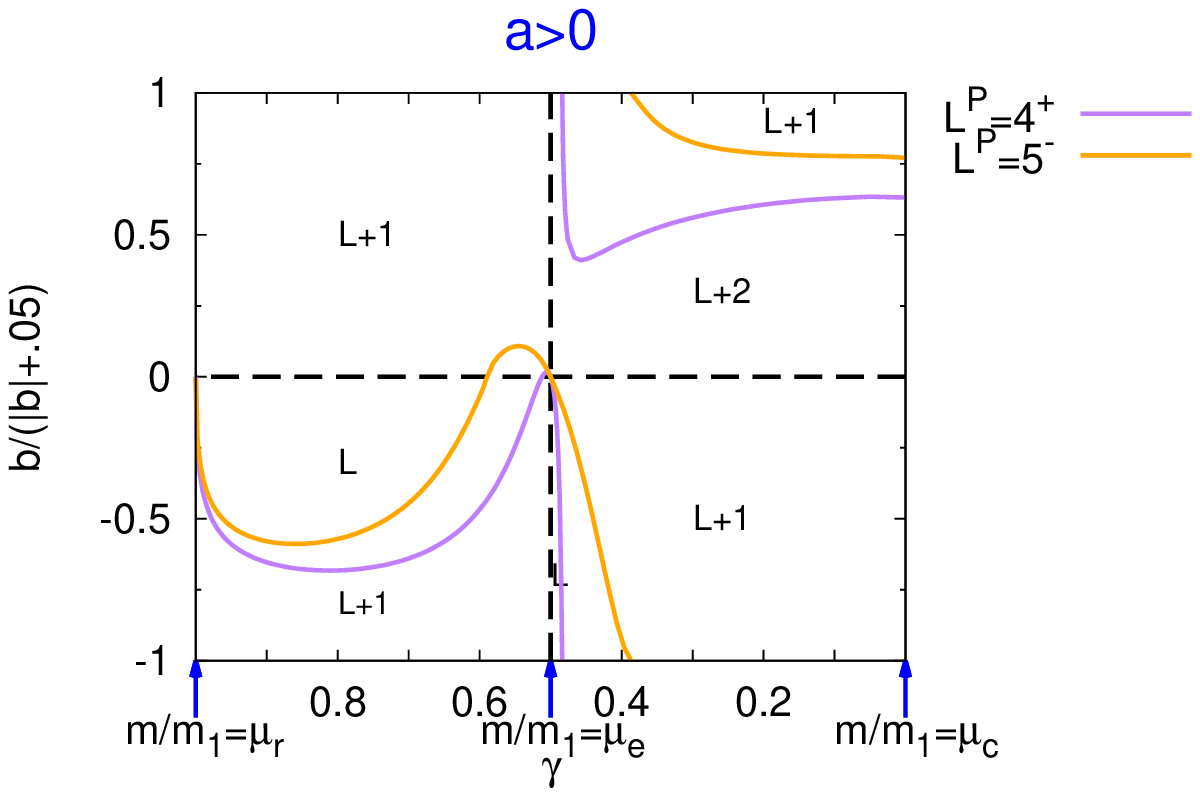}\hspace{-.7cm}\\
\includegraphics[width=.52\textwidth]{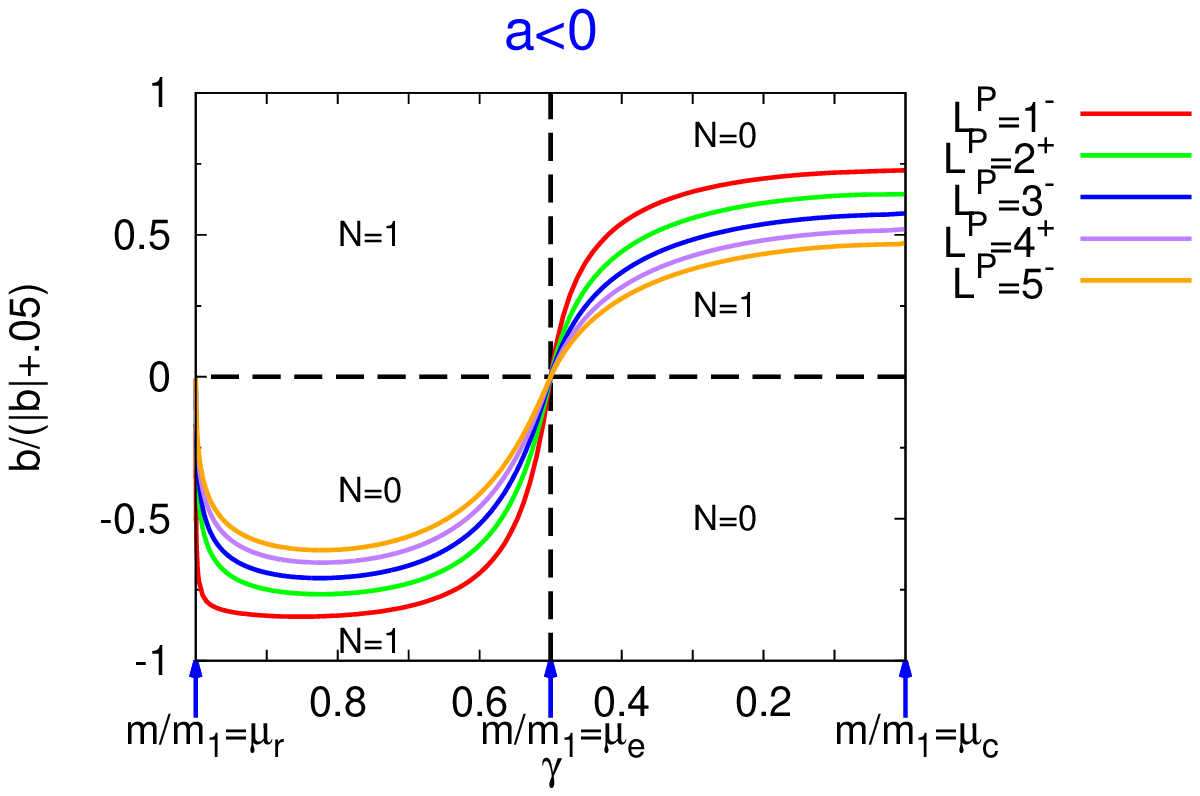}\hspace{-.7cm}
{\caption{A number of bound states in $L^{P}$ sector in each domain 
of the $\gamma$ -- $b$ plane for the two-body scattering length $a > 0$ 
and  $a < 0$. 
Solid line: critical three-body parameter $b_c(\gamma)$, for which 
the bound-state energy coincides with the threshold. 
Dashed (black) lines: domain boundaries determined by $\gamma = 1/2$ 
($m/m_1 = \mu_e$) and  $b = 0$. 
Values $\gamma = 1$, $1/2$ and $0$ correspond to $\mu_r$, $\mu_e$ and 
$\mu_c$. 
} 
\label{fig_mub}}
\end{figure*}

\begin{table}[htb]
\caption{\small Bound-state energies ($E_i$) for $b = 0$ and 
$b \to \infty$  at the critical values $m/m_1\to\mu_{r}$, 
$m/m_1 = \mu_{e}$ and $m/m_1 = \mu_{c}$ and $a>0$. 
The entry for $E_0$ and  $m/m_1\to\mu_{r}$ represent also 
the energies in the case $a<0$.
Two identical particles are fermions (bosons) if parity is odd (even).} 
\label{tab2}
\begin{tabular}{cccccc}
$L^P$ & $1^-$ & $2^+$ & $3^-$ & $4^+$ & $5^-$ \\
  \cline{2-6}
 &&& $m/m_1 \to \mu_r$ &&\\
 \cline{2-6}
$-E_0$ & 4.7473  & 11.3111  &  21.1146 &  34.1622 & 50.4592   \\
$-E_1$ & 1.02090 & 1.68551  &  2.77004 &  4.22117 &  6.03404  \\
$-E_2$ & -       & 1.02748  &  1.35435 &  1.85585 &  2.49935  \\
$-E_3$ & -       & -        &  1.03169 &  1.24191 &  1.54982  \\
$-E_4$ & -       & -        & -        &  1.03374 &  1.18686  \\
$-E_5$ & -       & -        & -        &  -       &  1.03485  \\
 \cline{2-6}
&&& $m/m_1 = \mu_e$ &&\\
 \cline{2-6}
$-E_0$ & 1.74397 & 3.42540  &  5.90130 & 9.17834  & 13.26233   \\
$-E_1$ & -       & 1.27038  &  1.86005 & 2.66958  & 3.68670   \\
$-E_2$ & -       & -        &  1.16759 & 1.49716  & 1.93501   \\
$-E_3$ & -       & -        & -        & 1.12596  & 1.34729   \\
$-E_4$ & -       & -        & -        & 1.00088  & 1.10398  \\
$-E_5$ & -       & -        & -        & -        & 1.00503  \\ 
  \cline{2-6}
&&& $m/m_1 = \mu_c$ &&\\
 \cline{2-6}
$-E_0$ & 5.89543 & 12.67370 & 22.57676 & 35.67806 & 52.00787  \\
$-E_1$ & 1.13767 &  1.84445 &  2.93742 &  4.39267 &  6.20802  \\
$-E_2$ & -       &  1.07220 &  1.41376 &  1.91816 &  2.56267  \\
$-E_3$ & -       & -        &  1.05497 &  1.27207 &  1.58177  \\
$-E_4$ & -       & -        & -        &  1.04795 &  1.20485  \\
$-E_5$ & -       & -        & -        & -        &  1.04442  \\ 
 \cline{2-6}
\end{tabular}
\end{table} 
\subsection{Critical conditions}
\label{critical_conditions}
\subsubsection*{Dependence of the number of bound states on the mass ratio 
and the three-body parameter}

For better illustration of the energy spectrum it is useful to construct 
the ''Phase Diagram'' representing appearance of the bound state in the 
plane of two parameters $m/m_1$ and $b$. 
In this respect it is necessary to take into account the following. 
If $b$ passes over $b = 0$ from negative to positive values, one 
additional bound state arises at $-\infty$ and it's energy increases with 
increasing $b$,~i.e., a number of bound states $N(b)$ increases by unity 
for $b > 0$ ($N(b \geq 0) = N(b < 0) + 1$). 
Furthermore, if the mass ratio goes across $\mu_e$ from lower to higher 
values and $b \neq 0$, one more bound state should appear for $a > 0$ and 
disappear for $a < 0$. 
If $b$ passes critical line $b_c(m/m_1)$ (for which the bound-state energy 
$E$ coincides with the threshold) from higher to lower values of $b$ one 
more bound state should appear both for positive and negative value of 
$a$. 
The critical line $b_c(m/m_1)$ goes from the point $b_c=0$, $m/m_1=\mu_r$  
through the point $b_c=0$, $m/m_1=\mu_e$ to the point 
$m/m_1 = \mu_c$, $b_c = b_f > 0$. 
Therefore, the lines $b_c(m/m_1)$, $b = 0$, and $m/m_1 = \mu_e$ form 
boundaries of the domains of the definite number of bound states in 
the $m/m_1$ -- $b$ plane as presented in Fig.~\ref{fig_mub}. 

Elaborate calculations were carried out to determine the critical 
parameter $b_c(m/m_1)$, for which the bound-state energy coincides with 
the threshold. 
Namely, $b_c(m/m_1)$ was determined by solving the eigenvalue problem for 
HREs at the two-body threshold $E = -1$ for the two-body scattering length 
$a > 0$ and at the three-body threshold $E = 0$ for $a < 0$. 
Existence of bound states at the threshold energy follows from 
the power decay of the channel function $f_1(\rho)$ for $\rho \to \infty$, 
namely, $f_1(\rho) \sim \rho^{-L}$ for $a > 0$ and 
$f_1(\rho) \sim \rho^{ - L - 3/2}$ for $a < 0$, that is related with 
asymptotic behaviour of the first channel effective 
potential~(\ref{V1apos_inf}) and~(\ref{Vnaneg_inf}) for positive and 
negative $a$, respectively. 
As usual, the bound state at the threshold turns to a narrow resonance 
under small variations of $m/m_1$ and $b$.

Few points of the dependence $b_c(m/m_1)$ are of special interest,~viz., 
one finds for $a > 0$ that $b_c = 0$ at largest mass ratios presented in 
upper part of Tab.~\ref{tab3} (namely, $m/m_1 \approx 12.91742$, 
$37.7662$, $74.8233$, $124.168$, $185.829$ for 
$L^P = 1^-,\,2^+,\, 3^-,\, 4^+,\, 5^-$, respectively), 
$b_c \to \pm \infty$ at mass ratios presented in lower part 
of Tab.~\ref{tab3};  $b_c \approx 0.0517$, $0.0416$, $0.0547$, $0.0897$, 
$ 0.177$ for $1^-,\,2^+,\, 3^-,\, 4^+,\, 5^-$, respectively, at the mass 
ratio $m/m_1 = \mu_c$; and $b_c(m/m_1)$ has a local minimum 
$b_c \approx -0.0175$ at $m/m_1 \approx 12.550$ for $L^P = 1^-$ sector, 
$b_c \approx -0.0095$ at $m/m_1 \approx 37.420$ for $L^P = 2^+$ sector, 
$b_c \approx -0.0038$ at $m/m_1 \approx 74.635$ for $L^P = 3^-$ sector, 
$b_c \approx -0.0009$ at $m/m_1 \approx 124.217$ for $L^P = 4^+$ sector, 
$b_c \approx -0.0062$ at $m/m_1 \approx 186.143$ for $L^P = 5^-$ sector. 
Similarly, one finds for $a < 0$ that $b_c \approx 0.13620$, $0.09065$, 
$0.06725$, $0.05324$, $0.04398$ for 
$L^P = 1^-,\,2^+,\, 3^-,\, 4^+,\, 5^-$, 
respectively, at $m/m_1 = \mu_c$; and $b_c(m/m_1)$ has a local minimum 
$b_c \approx -0.2501$ at $m/m_1 \approx 10.15$ for $L^P=1^-$, 
$b_c \approx -0.1634$ at $m/m_1 \approx 34.758$ for $L^P = 2^+$, 
$b_c \approx -0.1201$ at $m/m_1 \approx 71.980$ for $L^P = 3^-$, 
$b_c \approx -0.0947$ at $m/m_1 \approx 121.678$ for $L^P = 4^+$, 
$b_c \approx -0.0780$ at $m/m_1 \approx 183.837$ for $L^P = 5^-$. 
\begin{table}[htb]
\caption{The critical values $m/m_1$, for which the bound 
$L^P$ states arise for $b_c(m/m_1) = 0$ and $b_c(m/m_1) \to \infty$. 
Two identical particles are fermions (bosons) if parity is odd (even). 
} 
\label{tab3}
\begin{tabular}{cccccc}
$L^P$ & $1^-$ & $2^+$ & $3^-$ & $4^+$ & $5^-$  \\
 \cline{2-6}
& & & $b = 0$ && \\
 \cline{2-6}
 & 8.17259  & 22.6369 & 43.3951 & 70.457  & 103.823 \\
 &  12.91742 & 31.5226 & 56.1652 & 87.027  & 124.155 \\
 &  -        & 37.7662 & 67.3352 & 102.488 & 143.664 \\
 &  -        & -       & 74.8233 & 115.536 & 161.402 \\
 &  -        & -       & -       & 124.168 & 176.097 \\
 &  -        & -       & -       & -       & 185.829 \\
 \cline{2-6}
& & & $b \to \infty$ && \\
 \cline{2-6}
 & 10.2948  & 35.9163 & 73.9853 & 124.3660  & 187.056 \\
 \cline{2-6}
\end{tabular}
\end{table} 



\subsubsection*{Solution in the specific points $m/m_1 = \mu_e,\mu_c$}

A noticeable feature of the problem near $m/m_1 = \mu_e$ ($\gamma = 1/2$) 
is the degeneracy of energy dependences for different $b$ and a lack of 
continuity in the definition of $b$~(\ref{as_gam121}). 
It is not surprising as the sgn of the most singular term in HRE alters 
if $\gamma $ goes across $1/2$. 
Due to discontinuity in the definition of $b$ the limiting values of 
the bound-state energy for $m/m_1 \to \mu_e \mp 0$ 
($\gamma \to 1/2 \pm 0$) do not coincide with each other and with that 
calculated exactly at $m/m_1 = \mu_e$ ($\gamma = 1/2$). 
Notice also that in boundary condition~(\ref{as_gam121}) one could 
substitute $\log \rho$ with $\log (\rho/\rho_0)$ introducing a scale 
$\rho_0$, which simply leads to redefinition of length 
$\tilde {b} = b/(1 - b \log \rho_0)$. 

\begin{figure*}[htb]
\includegraphics[width=.45\textwidth]{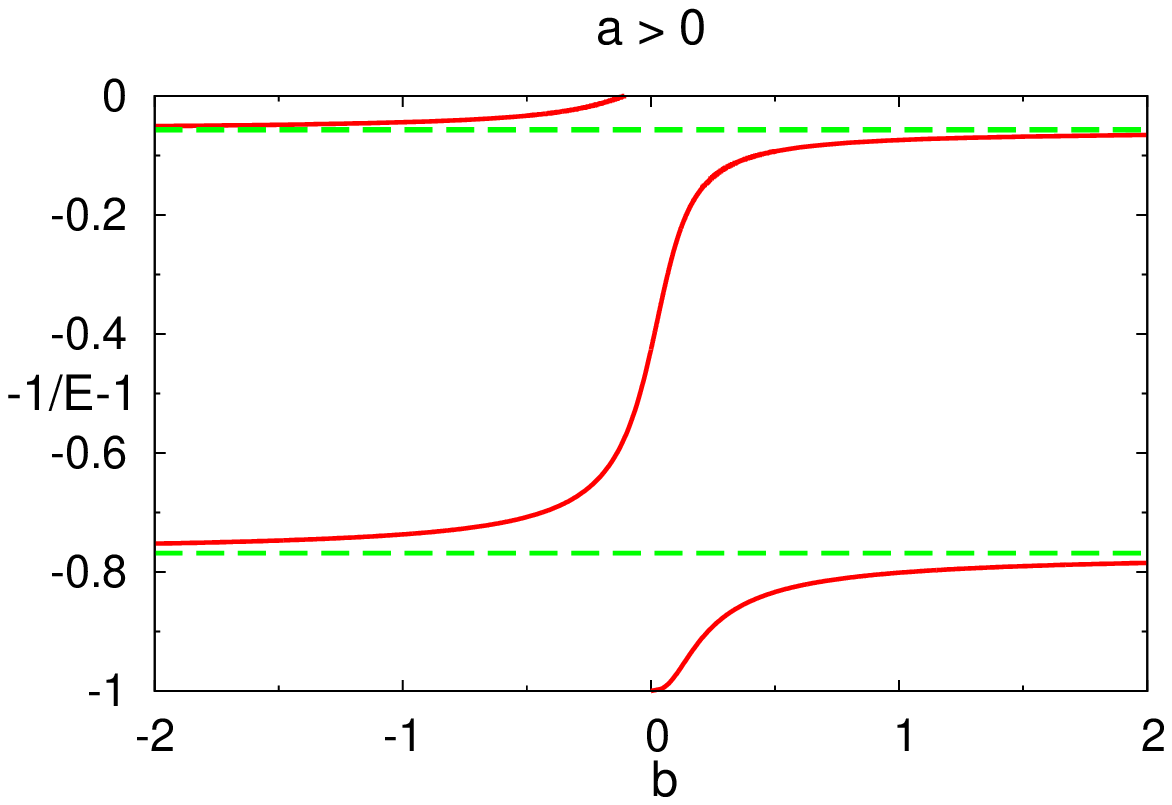}
\includegraphics[width=.45\textwidth]{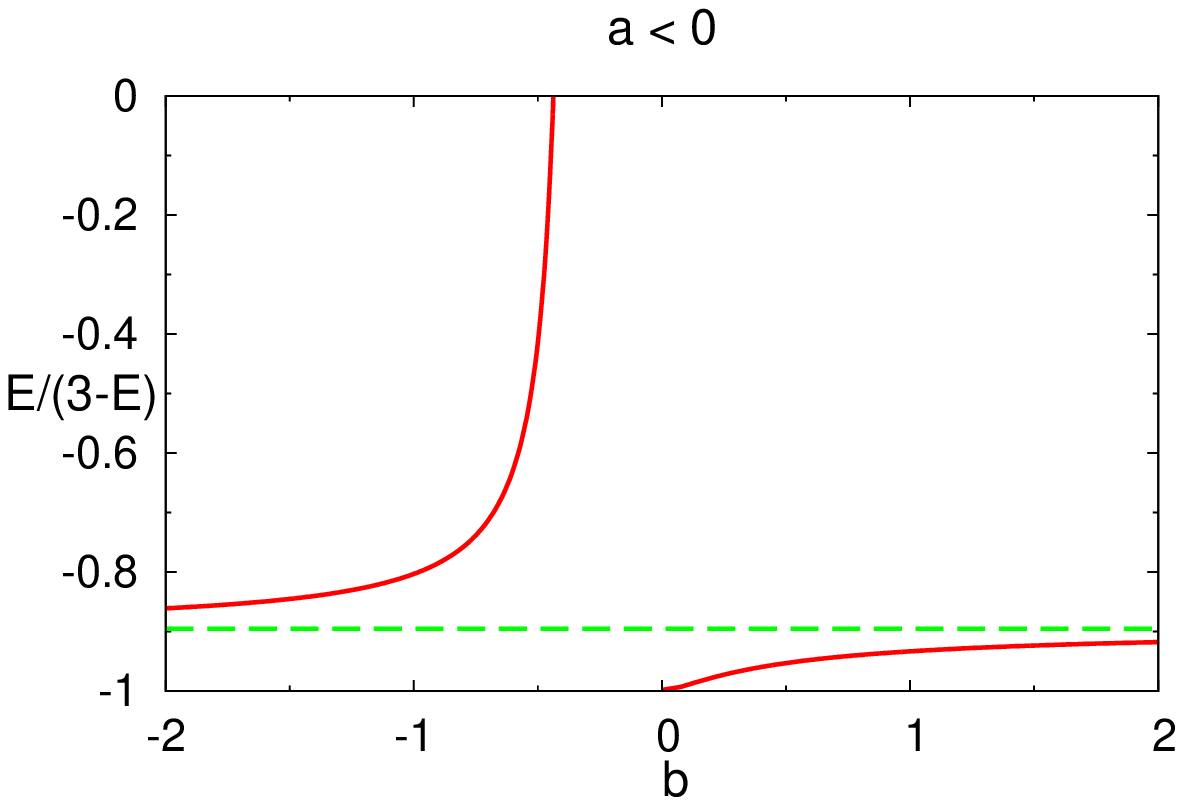}
{\caption{Bound-state energies $E$  for $L^P=1^-$ sector as a function of 
$b$ at $m/m_1 = \mu_e$ are plotted by solid (red) lines and asymptotic 
limits for $b \to \infty$ are indicated by dashed (green) lines. 
The two-body scattering length $a > 0$ (left) and $a < 0$ (right) and 
the energy axis scaled to map $-\infty < E < -1$ (left) and 
$-\infty < E < 0$ (right) to the interval $(-1, 0)$. 
} 
\label{Fig_mu_e} }
\end{figure*}

For illustration, the dependence of the bound-state energy on $b$ is 
calculated using boundary condition~(\ref{as_gam121}) and plotted in 
Fig.~\ref{Fig_mu_e} for $L^P=1^-$ sector. 
The calculations for $a > 0$ show that there are two bound states, one of 
which disappears for $-.108 < b \le 0$; for $a < 0$ there is one bound 
state, which disappears for $-.437 < b \le 0$. 
In the limit $b \to \infty$ the bound-state energies tend to $-4.319$ and 
$-1.061$ for $a > 0$ and to $-25.720 $ for $a < 0$. 
For $b = 0$ definitions~(\ref{as_gam121}) and~(\ref{as_gam121}) are the 
same and for $a > 0$ the bound-state energy takes the value 
$\sim -1.74397$. 




The bound-state energies and critical values of $m/m_1$ are in general 
agreement with results of~\cite{Kartavtsev07a,Endo11,Helfrich11} 
for $b = 0$. 
Only exception is that  the loosely bound $L^P = 3^-$ state is missed 
in calculation~\cite{Endo11}, which solves STM equation. 
The critical value $m/m_1$ of arising a ground $L^P=1^-$ state calculated  
in~\cite{Michelangeli13} is $8.1833$ that is close with $8.17259$.


\section{Discussion and conclusion}

One of the essential points of the three-body problem under consideration 
is regularization, which is necessary for some values of the mass ratios. 
Under regularization, one introduces an additional parameter describing 
the wave function in the vicinity of the triple collision point. 
The aim of the present paper is to describe the problem for the mass-ratio 
interval $\mu_r < m/m_1 \le \mu_c$, where $\mu_c$ is a critical value, 
above which the ''falling-to-center" or Efimov and Thomas effects take 
place. 
The principal part of the problem could be considered in terms of 
the singular interaction $\frac{\gamma^2-1/4}{\rho^2}$, where the interval 
$0 \le \gamma^2 < 1$ corresponds to $\mu_c \ge  m/m_1 > \mu_r$ and for 
$\gamma^2 < 0$, corresponding to $ m/m_1 > \mu_c$, the spectrum is not 
bounded below. 
Introducing an additional parameter $b$ one selects the unique solution 
in the limit $\rho \to 0$. 
It is necessary to emphasize that to introduce the parameter $b$ for 
$1/4\ge \gamma^2 < 1$  one should consider also the less singular term 
$\sim \frac{q}{\rho}$ of the interaction. 

Both total angular momentum and parity are good quantum numbers that 
allows one to provide a description of full spectrum by calculating 
the bound states separately in different $L^P$ sectors. 
If two identical particles are fermions (bosons) the bound state exists
only in odd (even) states. 
Dependence of bound-state energies on  $b$ and $m/m_1$ are calculated 
for some of $L^P$ sectors, a number of the bound states increases with 
increasing $L$ for two-body scattering length $a>0$. 
For negative two-body scattering length at most one bound state exists. 

It is of interest to find if the described above scenario happens 
for the three-body problem in the mixed 
dimensions~\cite{Nishida08a,Nishida11,Lamporesi10}, in the presence of 
spin-orbit interaction~\cite{Shi14,Cui14,Shi15} and on the lattice. 
The disclosed dependence on the three-body parameter should be taken into 
account to study of many-body properties as well,~e.g., in the four-body 
($3 + 1$)~\cite{Castin10} and ($2 + 2$)~\cite{Endo15a} problems. 
Up to now there are calculations, which show that the critical value of 
the mass ratio, above which the spectrum is not bounded below, are 
$m/m_1 \approx 13.607$~\cite{Efimov73} for $2 + 1$ problem ,
$m/m_1 \approx 13.384$~\cite{Castin10} for $3 + 1$ problem and 
$m/m_1 \approx 13.279$~\cite{Bazak17} for $4 + 1$ problem. 
The first bound state is known to appear at 
$m/m_1 \approx 8.17259$~\cite{Kartavtsev07} for $2 + 1$ problem,
at $m/m_1 \gtrsim 8.862$~\cite{Bazak17} for $3 + 1$ problem and
at $m/m_1 \gtrsim 9.672$~\cite{Bazak17} for $4 + 1$ problem. 
Concerning the four-body ($2 + 2$) problem of two fermions of one 
species interacting with two fermions of another species, one
should mention that spectrum is bounded below, as stated in 
papers~\cite{Endo15a,Michelangeli16}, and the proof of this statement 
for mass ratio in the interval $[0.58 , 1.73]$ is given in~\cite{Moser18}. 
For $N$ identical fermions interacting with a distinct particle 
the spectrum is bounded below for 
$m/m_1 < (0.36)^{-1} \approx 2.778$~\cite{Moser17}.
There is another estimate in papers~\cite{Correggi12,Correggi15a}, which 
give $m/m_1 \approx 5.291$ for $N = 3$, $\ m/m_1 \approx 1.056$ for 
$N = 8$, $\ m/m_1 \approx .823$ for $N = 9$. 

Furthermore, it is of interest to study $m/m_1$ and $b$ dependencies of 
the scattering cross sections, including the three-body resonances and 
the recombination rate. 
Another point is to consider a role of the three-body parameter $b$ for 
$m/m_1$ near $\mu_c$~\cite{Endo12,Gao15} and for $m/m_1$ near $\mu_r$. 

\bibliography{fermions,ferm_zrm,ferm_footnote}

\appendix

\section{Solutions of the auxiliary problem on a hyper-sphere}
\label{Appendix_hypsol}

The unnormalized solutions of the equation~(\ref{eqonhyp1}) 
and the boundary condition~(\ref{bconhyp1}) is given by 
\begin{eqnarray}
\nonumber
\hskip -.3cm  \varphi^L(\alpha ) && \sim \sqrt{\sin\alpha} 
Q_{\gamma - 1/2}^{L + 1/2} (\cos\alpha ) = 
 \sin\frac{\alpha}{2} \left( \tan \frac{\alpha}{2}\right)^{L} 
\frac{\Gamma(1 + \gamma + L)\,
\Gamma(-L-1/2)}
{2 \ \Gamma(\gamma - L)}\times\\
\ &&\times 
\ F\left(\frac{1}{2} - \gamma, \frac{1}{2} + \gamma; L + \frac{3}{2};
\sin^2\frac{\alpha}{2}\right) \ , 
\label{varphi_qf}
\end{eqnarray}
where $Q_\nu^\mu(x)$ and $F(a,b;c;x)$ are the Legendre function of 
second kind and hypergeometric function~\cite{Bateman53}. 
Another form 
\begin{eqnarray}
\nonumber
\hskip -.3cm  \varphi^L(\alpha ) && \sim (\sin\alpha)^{L+1}
\ F \left(\frac{L+\gamma+1}{2}, \frac{L-\gamma+1}{2} ; L + \frac{3}{2};
\sin^2\alpha\right) \, 
\label{varphi_qf1}
\end{eqnarray}
has been used in~\cite{Castin11,Gasaneo02}. 
In fact, it is a finite sum, which can be written as 
\begin{equation}
\label{varphi}
 \varphi^L(\alpha ) = \left(\sin\alpha\right)^{L + 1}
\left(-\frac{d}{d\cos\alpha}\right)^L
\frac{\sin\gamma\alpha}{\sin\alpha}\, .
\end{equation}
In the limit $\gamma \to n$ Eq.~(\ref{varphi}) reduces
\begin{equation}\label{varphi_pol}
 \varphi^L(\alpha ) \to \left(\sin\alpha\right)^{L + 1}
\left(-\frac{d}{d\cos\alpha}\right)^L
U_{n-1}(\cos\alpha)\ 
\end{equation}
via the Chebyshev polynomial $U_n(x)$, while for $\gamma = 1/2$ Eq.~(\ref{varphi}) 
simplifies to 
\begin{equation}
\label{varphi_g12}
 \varphi^L(\alpha ) \sim 
\sin\frac{\alpha}{2}\left(\tan\frac{\alpha}{2}\right)^L\, .
\end{equation} 

To derive the transcendental equation of the form~(\ref{transeq1}), one should 
use both $\varphi^L(\alpha, \rho)$ and its derivative for 
$\alpha = \pi/2$, 
\begin{eqnarray}
\label{LegQ0}
&&  \varphi^L(\pi/2)  = \frac{2^{L - 1}\sin \pi\gamma}{\pi}
\Gamma\left(\frac{L + \gamma + 1}{2}\right)
\Gamma\left(\frac{L - \gamma + 1}{2}\right)\, , \\ 
\label{varphi_deriv_pi/2}
&& \frac{d  \varphi^L(\alpha )}{d \alpha} 
\Big|_{\alpha = \pi/2} =  \varphi^{L+1}(\pi/2)\ . 
\end{eqnarray} 

Alternatively, one can use the recurrent relations for the Legendre
functions~\cite{Bateman53} to express 
\begin{equation}
\label{varphi_AB}
\varphi^L(\alpha ) = \displaystyle \left[
A_{L, \gamma}(\cot\alpha)\sin\gamma\alpha + 
B_{L, \gamma}(\cot\alpha)\cos\gamma\alpha\right] \ , 
\end{equation}
where 
\begin{subequations}
\label{AB}
\begin{equation}
A_{L, \gamma}(x) = (\gamma - L)x A_{L - 1, \gamma}(x) - (\gamma + L - 1)
\left[ x A_{L - 1, \gamma - 1}(x) + B_{L - 1, \gamma - 1}(x)\right], 
\end{equation}
\begin{equation}
B_{L, \gamma}(x) = (\gamma - L)x B_{L - 1, \gamma}(x) + (\gamma + L - 1)
\left[ A_{L - 1, \gamma - 1}(x)- x B_{L - 1, \gamma - 1}(x)\right] , 
\end{equation}
\end{subequations}
satisfy the recurrent relations, which start from $A_{0, \gamma}(x) = 1$, 
$B_{0,\gamma}(x) = 0$.
Few lowest-$L$ coefficients are $A_{1, \gamma}(x) = -x$,
$B_{1,\gamma}(x) = \gamma$, $A_{2, \gamma}(x) = 1 - \gamma^2 + 3x^2$,
$B_{2,\gamma}(x) = -3\gamma x$, $A_{3, \gamma}(x) = 3x (2\gamma^2 - 3 
- 5x^2)$, $B_{3,\gamma}(x) = \gamma (15x^2 + 4 - \gamma^2)$. 

\subsection{Leading order terms in the small-hyperradious expansion for large angular momentum $L$}

For large values of $L$ one obtains $\tilde\gamma$ and $q$ using few 
terms of the expansion of hypergeometric function in~(\ref{varphi_qf}).
After substitution of the expansion in the transcendental 
equation~(\ref{transeq1}), one obtains for $\rho=0$ up to $(L+1/2)^{-7}$
\begin{equation}\label{Ul}
\cos\omega \approx \frac{u_0}{L+1/2} +\frac{u_1}{(L+1/2)^3} + \frac{u_2}{(L+1/2)^5}, 
\end{equation} 
where $u_i$ are defined as 
\begin{subequations}\label{ui}
\begin{eqnarray}
 \hskip -12.4cm 
 u_0  = e^{-u_0}\ ,
\end{eqnarray}
\begin{eqnarray}\hskip -7.1cm  \label{u1}
&& u_1  = u_0\left( \frac{\tilde\gamma^2 - \frac{1}{4}}{2}
+ \frac{u_0^3}{1 + u_0} \left(\frac{1}{4} - \frac{u_0}{3}\right)
\right) , 
\end{eqnarray}
\begin{eqnarray}\nonumber
u_2 =\frac{3 \tilde\gamma^4}{8}&&-\frac{21(1+u_0)- 6 u_0^2 +
 32 u_0^3}{48(1 + u_0)}\tilde\gamma^2+\\
&&\frac{495 + 1485 u_0 + 1305 u_0^2 + 1095 u_0^3 + 3360 u_0^4 - 
912 u_0^5 - 1744 u_0^6 + 768 u_0^7}{5760(1 + u_0)^3} ,
\end{eqnarray}
\end{subequations}
which approximately gives $u_0 \approx 0.567143$,
$u_1\approx -0.0637978 + 0.283572\, \tilde\gamma^2$, 
$u_2\approx 0.056468 - 0.277587\, \tilde\gamma^2 + 0.212679\, 
\tilde\gamma^4$. 
As $\cos\omega=\frac{\sqrt{1+2m/m_1}}{1+m/m_1}$, the connection 
of the critical mass ratios can be founded 
\begin{equation} 
\label{mucr_cer}
m/m_1 \approx \frac{2}{u_0^2}\left(l^2 - \tilde \gamma^2\right) + 
\upsilon_0 + \frac{\upsilon_1 (\tilde \gamma^2)}{l^2}+O(l^{-4}) \ , 
\end{equation}
where 
\begin{subequations}\label{vi}
\begin{equation} 
\hskip -8.4cm
 \upsilon_0= \displaystyle \frac{1}{2u_0^2} - \frac{5/2 + u_0/6}
{1 + u_0} \ ,
\end{equation}
\begin{eqnarray}
\nonumber
\upsilon_1 (\tilde \gamma^2) = \tilde\gamma^2 && 
\left(\frac{1}{u_0^2} + 
 \frac{1 + 2 u_0/3}{1+u_0}\right) - \\
&&\frac{90 + 270 u_0 + 360 u_0^2 + 330 u_0^3 + 525 u_0^4 -
 48 u_0^5 - 181 u_0^6 - 3 u_0^7}{360 u_0^2 (1 + u_0)^3}  
\end{eqnarray}
\end{subequations}
and $\frac{2}{u_0^2} \approx 6.2179$, 
$\upsilon_0 \approx -.101098$ and 
$\upsilon_1 \approx -1.04234 + 3.98832\, \tilde{\gamma}^2$. 
The terms of Eq.~(\ref{mucr_cer}) up to a constant coincide with presented 
in~\cite{Castin11}. 
Comparing these values with those given in Table~\ref{tab1},
reveals that relative accuracy is better then $10^{-4}$ for $L=5$ 
and $10^{-5}$ for $L=10$.
The relationship $\mu_c - \mu_r = 4 (\mu_c - \mu_e) + O((L+1/2)^{-2})$ 
also follows from Eq.~(\ref{mucr_cer}) for large $L$. 

The dependence $\tilde{\gamma}^2 (m/m_1)$ up to an order 
$O((L + 1/2)^{-2})$ is a linear function presented in Fig.~\ref{fig_gamma}. 
In a similar way, $q(m/m_1)$ up to an order $O((L+1/2)^{-2}) $ is 
a square-root dependence presented in Fig.~\ref{fig_gamma} as
\begin{equation}\label{qas}
q \approx - \frac{2 u_0 \sign(a) }{(1 + u_0) \cos\omega}
 \left( 1 +\frac{c_q}{l^2} + O(l^{-4} )\right)\ ,
\end{equation} 
where $\frac{2u_0}{(1+u_0)}\approx 0.7237925$, 
$c_q=\frac{1}{2}+\frac{u_0^2(3+2u_0)}{6(1+u_0)}\approx 0.641425$.
Accuracy of~(\ref{qas}) is about $10^{-3}$ for $L=5$ and $4 \cdot 10^{-5}$ for $L=10$. 

\section{Zero-range limit of the three-body potential}

\label{Appendix_bc}
In relation with the discussion in Sec~\ref{TCP}, it is of interest to 
analyse zero-range model in the presence of general centrifugal and 
Coulomb interaction, namely, to consider the Schr\"{o}dinger equation for 
$0 \leq \gamma < 1$
\begin{equation}
\label{shred_shrink}
\displaystyle \left[-\frac{d^2}{d \rho^2} + \frac{\gamma^2 - 1/4}{\rho^2} 
+ \frac{q}{\rho} +
\frac{\lambda}{\rho_0^2} V(\rho/\rho_0)- E \right] f(\rho) = 0\, ,  
\end{equation}
in the limit $\rho_0 \to 0$. 
In this limit, a shape of the short-range potential $V(x)$ becomes 
insignificant and one comes to one-parameter description of solutions. 
As in Sec.~\ref{TCP}, it is natural to use the generalized scattering 
length $b$ defined by Eqs.~(\ref{as_gam121})--(\ref{as_gam12}) as 
a parameter.
One expect that for any dependence $\lambda(\rho_0)$ the GSL $b$ is 
determined by the limit
\begin{equation}
\label{blam}
b \xrightarrow[\rho_0 \to 0]{} A \, \rho_0 \,    
\frac{\sign\left(\lambda - \lambda_c\right)}
{\left| \lambda - \lambda_c - B\, q\, \rho_0\right|^{\frac{1}{2\gamma}}}
\ ,
\end{equation}
except $\gamma = 0, \ 1/2$.
The constants $\lambda_c$, $A$, $B$ are specified by $\gamma$ and a form 
of the potential $V(\rho/\rho_0)$. 
The values $\lambda_c$ have a meaning of critical values of the strength 
of potential, at which the threshold bound state appears. 
Note that, as in Section~\ref{TCP}, the  Coulomb interaction plays no role 
in definition of $b$ for $\gamma < 1/2$, therefore, for this interval 
Eq.~(\ref{blam}) reduces to the simpler expression with $B=0$.  
The parameter $A$ is not crucial due to it can be included in definition 
of $b$. 

In the limit $\gamma = 0$, only possible positive values of $b$ 
are determined by
\begin{equation}
\label{blam_0}
b \xrightarrow[\rho_0 \to 0]{} A \, \rho_0 \,    
\exp \left(\frac{1}{\lambda - \lambda_c}\right)\ . 
\end{equation}
In the special case $\gamma = 1/2$, 
\begin{equation}
\label{blam_12}
b \xrightarrow[\rho_0 \to 0]{} A \, \rho_0 \,    
\left[\lambda - \lambda_c - B\, q\, \rho_0 \log \rho_0\right]^{-1}\ ,
\end{equation}
i.~e.,  $b$ is the usual scattering length if $q = 0$ and the Coulomb 
modified scattering length if $q \neq 0$. 

From Eqs.~(\ref{blam})--~(\ref{blam_12}), it is clear that $b = 0$ for 
any limiting values $\displaystyle\lim_{\rho_0 \to 0}\lambda(\rho_0)$, except 
for $\lambda \xrightarrow[\rho_0 \to 0]{} \lambda_c$, which correspond to 
existence of the bound or virtual state at zero energy. 
Any values of $0 < |b| < \infty$ is determined by the dependence 
$\lambda(\rho_0)$ in the vicinity of $\lambda_c$. 
In other words, for finite (infinite) $b$, the dependence on the potential
range should be of the form 
\begin{equation}
\label{lam}
\lambda \xrightarrow[\rho_0 \to 0]{} \lambda_c + 
B \, q\, \rho_0 + \sign(b) \left(A\frac{\rho_0}
{|b|}\right)^{2\gamma} 
\end{equation}
where $\gamma \neq 1/2$; the term proportional to $q \rho_0$ can be omitted 
for $0 < \gamma < 1/2 $.  
In the limit $\gamma \to 0$, Eq.~(\ref{lam}) reduces to
\begin{equation}
\label{lam_0}
\lambda \xrightarrow[\rho_0 \to 0]{} \lambda_c + 
\log \left(A\frac{ \,\rho_0} {b}\right)\ . 
\end{equation}
In the special case $\gamma = 1/2$, 
\begin{equation}
\label{lam_12}
\lambda \xrightarrow[\rho_0 \to 0]{} \lambda_c + 
B \, q\, \rho_0 \log \rho_0 + A\frac{\rho_0}{b}  \ .
\end{equation}
One should underline that limit $\lambda \to \lambda_c$, mentioned
in literature as "resonance" condition, corresponds not only to
$b \to \infty$, rather to $b \neq 0$.

\subsection*{Lennard-Jones and similar potentials}

For an illustration of the above general considerations, one can use 
in Eq.~(\ref{shred_shrink}) a class of Lennard-Jones LJ $(m,n)$ 
potentials of the form $V(x) = \left[x^{-m} - x^{-n}\right]$, which is 
common for the inter-atomic interactions and applied to the three-fermion 
problem in~\cite{Nishida08}. 

In particular, the analytical zero-energy solution of~(\ref{shred_shrink}) 
can be obtained if $q = 0$ for LJ $(2n + 2, n + 2)$ potentials with restriction $n > 2 $, namely,  
\begin{equation}
\label{fLG}
f(x) = \displaystyle \sqrt{x}\, e^{-\frac {\sqrt{\lambda}} {n x^n}} 
\left[ \,x^\gamma 
\,\Phi \left( \frac{1}{2} - \frac{\sqrt{\lambda}}{2n} - \frac{\gamma}{n}, 
1 - \frac{2\gamma}{n}; \frac{2\sqrt{\lambda}}{n x^n}\right) - 
C \, x^{-\gamma} \,\Phi \left( \frac{1}{2} - \frac{\sqrt{\lambda}}{2n} + 
\frac{\gamma}{n}, 1 + \frac{2\gamma}{n}; 
\frac{2\sqrt{\lambda}}{n x^n}\right) \right]\, , 
\end{equation}
where $\displaystyle \Phi(a,b;z)$ is the confluent hyper-geometric 
function and the coefficient  
\begin{equation}
\label{c2c1}
C = \,\displaystyle\frac { \Gamma(1 - \frac{2\gamma}{n}) 
\Gamma(\frac{1}{2} - \frac{\sqrt{\lambda}}{2n} + \frac{\gamma}{n})}
{\Gamma(1 + \frac{2\gamma}{n}) 
\Gamma(\frac{1}{2} - \frac{\sqrt{\lambda}}{2n} - \frac{\gamma}{n})}
\left(\frac{2\sqrt{\lambda}}{n }\right)^\frac{2\gamma}{n}
\end{equation} 
is determined by the boundary condition $f \to 0$ at $x \to 0$ 
and asymptotic form $\displaystyle \Phi(a,b;z)\to \frac{\Gamma(b)}
{\Gamma(a)} e^{z} z^{a - b}(1 + O(z^{-1}))$ for $|z| \to \infty$. 
By taking into account Eq.~(\ref{c2c1}) and comparing Eq.~(\ref{fLG}) 
for $\rho_0 \to 0$ ($x \to \infty$) with Eq.~(\ref{as_gam121}), one 
comes to
\begin{equation}
\label{bLJ126}
b \to \, S_\lambda\ \rho_0
\left|
\frac{ \Gamma (1 - \frac{2\gamma}{n}) 
 \Gamma(\frac{1}{2} - \frac{\sqrt{\lambda}}{2n} + \frac{\gamma}{n})}
 {\Gamma(1 + \frac{2\gamma}{n}) 
 \Gamma(\frac{1}{2} - \frac{\sqrt{\lambda}}{2n} - \frac{\gamma}{n})}
\right|^{\frac{1}{2\gamma}}
\left(\frac{2\sqrt{\lambda}}{n}\right)^\frac{1}{n}\,
\end{equation} 
where $S_\lambda=-1$ for $\sqrt{\lambda_{c}(N)} - 4\gamma < 
\sqrt\lambda < \sqrt{\lambda_{c}(N)}$, and $S_\lambda=1$ otherwise. 
The critical interaction strength  $\lambda_c$ for the $N$th state 
($N=0,1,2,...$) equals 
\begin{equation}
\label{lambda_LJ}
\sqrt{\lambda_{c}(N)} = 2\gamma + n + 2 n N\ 
\end{equation}
and corresponds to the infinite scattering length ($b \to \infty$).
In the limit $\lambda \to \lambda_c $ the generalized scattering length 
$b$ confirm the form Eq.~(\ref{blam}) if only $q = 0$, where 
\begin{eqnarray}
A = \left(\frac{2\sqrt{\lambda_c}}{n }\right)^\frac{1}{n} \left|
\frac{4n\sqrt{\lambda_c}} {\Gamma\left(\frac{2\gamma}{n}\right)} 
\right|^\frac{1}{2\gamma}\, . 
\end{eqnarray}
The scattering length~(\ref{bLJ126}) for the particular case 
$\gamma = 1/2$ coincides with~\cite{Pade07,Pade09}. 
For $1/2 < \gamma < 1 $ the term proportional to $q$ has to be taken 
into account, unfortunately, simple analytical expression~(\ref{lam}) 
is not obtained. 
For $\gamma= 0$ Eq.~(\ref{bLJ126}) reduces to
\begin{equation}
b = \, S_\lambda \ \rho_0 \left(\frac{2\sqrt{\lambda}}{n}\right)^\frac{1}{n}
\exp \left\{
\frac{1 }{n}\left(\psi\left(\frac{n-\sqrt{\lambda}}{2n}\right) 
+2 \gamma_c\right)
\right\}
\,
\end{equation}
that in the limit $\lambda \to \lambda_c = n (1 + 2N)$ confirm the 
dependence~(\ref{blam_0}) with
\begin{eqnarray}
A = \left(\frac{2\sqrt{\lambda_c}}{n }\ e^{\gamma_c}\right)^\frac{1}{n} \, . 
\end{eqnarray}

For LJ $(12,6)$ potential, as shown in~\cite{Nishida08}, the strength  
$\sqrt{\lambda} \to \sqrt{\lambda_c} \approx 
[6.5 + 2.9(\gamma - 1/2]^{5/6}$ 
corresponds to infinite scattering length $b \to \infty$. 
Varying $\lambda$ near $\lambda_c$ one can obtain any values of $b$.
More precise fit of numerical calculation of $\lambda_c$ gives 
$\sqrt{\lambda_c} \approx [6.460 + 2.903 (\gamma - 1/2]^{5/6}$. 
As the exact solution for LJ $(2n + 2, n + 2)$ potentials gives 
the linear dependence~(\ref{lambda_LJ}) for $\lambda_c$, the better 
fit $\sqrt{\lambda_c} \approx 4.729 + 1.773 (\gamma - 1/2)$ is expectedly 
obtained in numerical calculations. 
It is of interest to estimate the error of both fits by taking into 
account the next term, namely, $\sim(\gamma - 1/2)^2$.  
The coefficient in front of the next term is of order $0.1$ for dependence 
from~\cite{Nishida08} (moreover, it's modify the first constant from 
$6.460$ to $6.452$) and $-0.007$ for linear fit. 

{ Analytical solution at zero energy for $\gamma = l + 1/2$, where
$l = 0, 1, 2,...$, is known for potentials of Lennard-Jones type 
$(2n + 2, n + 2)$ with additional parameter $\tilde{\rho}$ that fix 
logarithmic derivative of inner part of the wave 
function~\cite{Szmytkowski95}. 
Procedure to calculate the scattering length was obtained for 
LG $(12, \{4,6,7\})$ ($\gamma = 1/2$) in~\cite{Gomez12}, application of 
LG $(2n + 2, n + 2)$ potentials ($\gamma = l + 1/2$, $l = 0, 1, 2,...$) 
was given in~\cite{Gao03} to calculate $Na$--$Na$ scattering $s$- and 
$d$-~wave cross sections.
Analytical solution at zero energy is known also for Lenz potentials 
($\gamma = l + 1/2$, $l = 0, 1, 2,...$)~\cite{Szmytkowski95}, 
potentials of polynomial, exponential type, Morse potential 
($\gamma = 1/2$)~\cite{Pade09}. }

\subsection*{Square-well potential}
The simple example is the potential defined as the square well 
\begin{equation}
U(\rho) = -\lambda \ \theta(\rho-\rho_0)/\rho_0^2 + 
\theta(\rho_0-\rho) \left(\displaystyle\frac{\gamma^2 - 1/4}{\rho^2} 
+ \frac{q}{\rho} \right)\, .
\end{equation}
The function $f(\rho ) = \sin \kappa \rho $ 
($\kappa \approx \frac{\sqrt{\lambda}}{\rho_0} $) for $\rho \le \rho_0$ 
and is of the form~(\ref{as_gam121}) for $\rho > \rho_0$ and 
$\gamma \neq 1/2$.
Matching the solutions at the point $\rho_0$ leads to
\begin{equation}
\label{kaprho}
\displaystyle \sqrt{\lambda}\cot\sqrt{\lambda} \approx 
\frac{\gamma + \frac{1}{2} \mp \left(\frac{|b|}{\rho_0}\right)^{2 \gamma }
\left(\frac{1}{2}-\gamma+\frac{q \rho_0}{1 - 2 \gamma}(\frac{3}{2}-\gamma) 
\right)}
{1\mp  
\left( \frac{|b|}{\rho_0}\right)^{2 \gamma }
\left(1+\frac{q \rho_0}{1 - 2 \gamma}\right)}\,  
\end{equation} 
that allow one to obtain 
\begin{equation}
\label{bsw}
b \xrightarrow[\rho_0 \to 0]{}
\rho_0\left[
\frac{\sqrt{\lambda} \cot {\sqrt{\lambda}}-\frac{1}{2}-\gamma}
{\sqrt{\lambda} \cot {\sqrt{\lambda}} 
\left(1+\frac{q\rho_0}{1-2\gamma}\right) + \gamma - 1/2 - 
\frac{q\rho_0}{1-2\gamma}\left(\frac{3}{2}-\gamma\right)}
\right]^{\frac{1}{2\gamma}}
\end{equation}
Expanding in Eq.~(\ref{bsw}) $\lambda$ near $\lambda_c$, determined by 
the lowest-value solution of $\sqrt{\lambda_c}\cot{\sqrt{\lambda_c}} = 
1/2 - \gamma$ , one comes to Eq.~(\ref{blam}) with 
\begin{equation}
\label{blamq}
A = \left[\frac{4\gamma\lambda_c}{\gamma^2 - 1/4 +
\lambda_c}\right]^{\frac{1}{2\gamma}}\ ,
\end{equation}
\begin{eqnarray}
\label{potrho}
B = \frac{\lambda_c}{(\gamma^2 - 1/4 + \lambda_c)(\gamma - 1/2)}  \ .
\end{eqnarray}
For $\gamma > 1/2$ the term proportional $B(\gamma)$ is 
important in contrast for $\gamma < 1/2$, where it can be omitted. 

Taking the limit $\gamma \to 0$, Eq.~(\ref{kaprho}) reduces to 
\begin{equation}
\label{kaprho_g0}
\displaystyle\sqrt{\lambda} \cot \sqrt{\lambda} 
\approx\ \frac{1}{2} + \frac{1}{\ln (\rho_0/b)}\, ,
\end{equation} 
where only $b > 0$ is allowed. 
Expending in equation 
\begin{equation}
\label{blam_g0}
b \xrightarrow[\rho_0 \to 0]{}  \rho_0 \, \exp\left[\frac{2}
{1 - 2 \sqrt{\lambda} \cot \sqrt{\lambda} } \right]\, 
\end{equation} 
following from Eq.~(\ref{kaprho_g0})), $\lambda$ near $\lambda_c$, determined by the lowest-value solution of 
$\sqrt{\lambda_c}\cot{\sqrt{\lambda_c}} = 1/2$ , one comes to 
Eq.~(\ref{blam_0}) with 
$A=\exp\left[\frac{8\lambda_c}{(4\lambda_c-1)} \right]$. 

For $\gamma = 1/2$, the function $f(\rho ) = \sin \kappa \rho $ 
($\kappa \approx \frac{\sqrt{\lambda}}{\rho_0} $) for $\rho \le \rho_0$ 
and is of the form~(\ref{as_gam12}) for $\rho > \rho_0$.
Matching the solutions at the point $\rho_0$ leads to
\begin{equation}
\label{kaprho_g12}
\displaystyle\sqrt{\lambda} \cot \sqrt{\lambda} 
\approx\ \rho_0\frac{1 - b q \left[1 +  \log\rho_0 \right]}{\rho_0 -
b(1 + q\rho_0\ln \rho_0)}\, .
\end{equation} 
Expending in equation 
\begin{equation}
\label{blam_g12}
b \xrightarrow[\rho_0 \to 0]{} \rho_0 \,
\frac{\sqrt{\lambda} \cot \sqrt{\lambda} - 1}
{\sqrt{\lambda} \cot \sqrt{\lambda} + q\rho_0\log\rho_0(\sqrt{\lambda}
\cot \sqrt{\lambda} - 1) - q \rho_0} \ ,
\end{equation} 
following from Eq.~(\ref{kaprho_g12})), $\lambda$ near 
$\lambda_c = \pi^2$, one comes to Eq.~(\ref{blam_12}) with $A=-2$, $B=-2$.

The equations analogous to Eqs.~(\ref{kaprho}) and~(\ref{kaprho_g0}) was 
obtained in~\cite{Bouaziz14} for $q=0$. 
In addition, $\delta$-shell regularisation was done in~\cite{Bouaziz14}.
Some discussion about square-well regularisation can be found 
in~\cite{Moroz15}.
For Efimov case, the square-well and $\delta$-shell regularizations 
was done in~\cite{Braaten04} for  $1/\rho^2$ potential. 


\subsection*{Two-parameter boundary condition}
\begin{figure*}[htb]
\includegraphics[width=.45\textwidth]{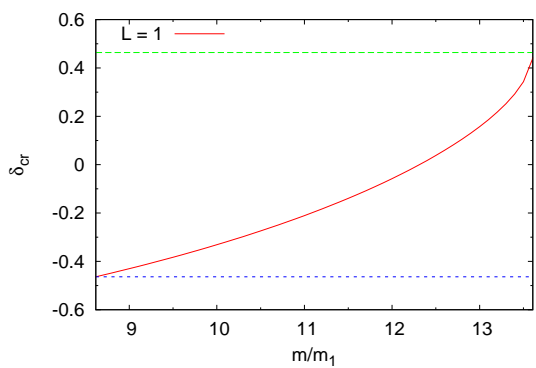}
{\caption{ $\delta_{cr}$. 
} 
\label{Fig_deltacr} }
\end{figure*}

Two-parameter boundary condition is introduced in 
paper~\cite{Safavi-Naini13}, by the logarithmic derivative 
of the function, $\tan\delta = \rho \displaystyle\frac{d \ln f}{d \rho}$
at small hyper-radius $\rho_0$.  
As follow from~(\ref{as_gam121}), in the limit $\rho_0 \to 0$ the three-body 
parameter $b$ is expressed via two parameters $\delta$ and $\rho_0$ as
\begin{equation}
\label{b_blume} 
|b|^{2\gamma} = 
\frac{\pm \rho_0^{2\gamma}\left[\tan\delta - \gamma - 
\frac{1}{2}\right]}{\left[1 + \frac{q \rho_0}{1 - 2\gamma}\right] 
\tan\delta + \gamma - \frac{1}{2} + q \rho_0\frac{2\gamma - 3}
{2(1 - 2\gamma)}} \ , 
\end{equation} 
except for $\gamma = 1/2$. 
Thus, $b$ is discontinuous at $\delta = \delta_{cr}$, where 
\begin{equation}
\label{tandc}
\tan\delta_{cr} = \frac{1}{2}-\gamma + 
\frac{q \rho_0}{1 - 2 \gamma + q \rho_0} +O(\rho_0^{2\gamma+1}).
\end{equation} 
For $\rho_0 \to 0$, Eq.~(\ref{tandc}) takes a simple form, 
$\tan\delta_{cr} = 1/2 - \gamma $ (the dependence $\delta_{cr}(m/m_1)$ 
is shown in Fig.~\ref{Fig_deltacr}), which is valid everywhere excluding 
a small neighbourhood $\sim q \rho_0$ of the point $m/m_1 = \mu_e$ (of 
the order of $|\gamma - 1/2| < q \rho_0$). 
To exemplify the correspondence between the model~\cite{Safavi-Naini13} 
and the present universal description, one compares $\delta_{cr}(m/m_1)$ 
obtained numerically in~\cite{Safavi-Naini13} and 
$\delta_{cr}(m/m_1) = \arctan (1/2 - \gamma )$.  
Comparison Fig.~5 of~\cite{Safavi-Naini13} and Fig.~\ref{Fig_deltacr} shows 
that $\delta_{cr}(m/m_1)$ are in agreement up to $m/m_1 \approx 13$, e.~g., 
$\delta_{cr} \to -\arctan(1/2) \approx -0.46$ for $m/m_1 \to \mu_r$ and 
$\delta_{cr} \to 0$ for $m/m_1 \to \mu_e$. 
On the other hand, the discrepancy arises above $m/m_1 \approx 13$, 
e.~g., the exact expression gives $\delta_{cr} \to \arctan(1/2) 
\approx 0.46$ for $m/m_1 \to \mu_c$, which differs from $\delta_{cr}$  
in fig.~5 of~\cite{Safavi-Naini13}. 
Presumably, this discrepancy indicates difficulty of the numerical 
calculation for $\rho_0 \to 0$ in this mass-ratio region. 

\end{document}